\documentclass[12pt]{article}

\newtheorem{theorem}{Theorem}
\newtheorem{proposition}{Proposition}  
  
\usepackage{vmargin}   
\setmarginsrb{2cm}{1cm}{2cm}{2cm}{1cm}{1cm}{1cm}{1.6cm}   
\usepackage{amsmath}   
\usepackage{amssymb}  
\usepackage[dvips]{graphicx}  
\usepackage{rotating}   
\usepackage{psfrag} 
\usepackage{fancyhdr}

\author{Karim Noui\thanks{email:noui@phys.univ-tours.fr} \\
Laboratoire de Math\'ematiques et de Physique Th\'eorique \\
F\'ed\'eration Denis Poisson \\
CNRS/UMR 6083, Facult\'e des Sciences et Techniques\\
Parc de Grandmont, 37200, Tours (EU)}

\title{\bf Three Dimensional Loop Quantum Gravity:\\
Particles and the Quantum Double}
\date{\today} 
 
\begin{document} 
\sloppy
\maketitle

\begin{abstract} 
It is well known that the quantum double structure plays an important role in three dimensional quantum gravity coupled to matter field.
In this paper, we show how this algebraic structure emerges in the context of three dimensional Riemannian loop quantum 
gravity (LQG) coupled to a finite number of massive spinless point particles. In LQG, physical states are usually 
constructed from the notion of $SU(2)$ cylindrical functions on a Riemann surface $\Sigma$ and the Hilbert structure is defined by the
Ashtekar-Lewandowski measure. In the case where $\Sigma$ is the sphere $S^2$, we show that the physical Hilbert space is in fact isomorphic to a 
tensor product of simple unitary representations of the Drinfeld double $DSU(2)$: the masses of the particles label the simple representations, 
the physical states are tensor products of vectors of simple representations and the physical scalar product is given by intertwining 
coefficients between simple representations. This result is generalized to the case of any Riemann surface $\Sigma$. 
\end{abstract} 

\newpage

\subsection*{1. Motivations}
Most of the articles dealing with any aspect of three dimensional gravity (3D) are introduced and motivated by the well-known fact 
that 3D gravity would provide a nice framework to answer some questions that are too cumbersome to solve in the 
four dimensional (4D) context. It is true that 3D (pure) gravity is a topological theory which admits only a finite
number of degrees of freedom: this makes the classical theory exactly solvable and different quantization schemes have been 
deeply studied in the literature \cite{Carlip}. The theory is nevertheless far from being trivial and exhibits very rich
mathematical structures that are still extensively studied: topological invariants, knots invariants, quantum groups, moduli spaces,
teichmuller spaces and so on... The main objection is that it is not true that all these techniques will shed light on the problem of
quantizing four dimensional general relativity. The interest of 3D gravity is widely limited by the fact that
there is no local degrees of
freedom in the theory and therefore there are many aspects of 4D gravity that will never be clarified thanks to this
toy-model. In order to make 3D gravity a richer tool, one should try to introduce local degrees of freedom in the theory.

Coupling 3D gravity to an external (non-trivial) field is a natural extension of the model that contains local degrees of freedom. 
Furthermore, it makes the model physically more interesting. But, the problem of defining a consistent (full) 
quantization of a self-gravitating field theory is very difficult and impossible to solve by making use of standard 
perturbative quantum field theory techniques, which are so successful for describing the physics of elementary particles.
Recently, Freidel and Livine \cite{Freidel} have proposed a solution of the problem in the context of 3D spin-foam models:
they have shown that (under certain hypothesis) a quantum self-gravitating field theory is equivalent to a non-self-gravitating 
but non-commutative quantum field theory. The authors claim that the non-commutativity encodes the quantum gravity effects 
on the field. Even if 3D gravity is physically irrelevant, it is rather interesting to have a model at hand where one has  
complete control of the quantum gravity effects. For these reasons, the article \cite{Freidel} certainly deserves attention
for it is a first step in the understanding of a theory of quantum gravity with local degrees of freedom. Since, other interesting
articles dealing with different aspects of this problem have been proposed \cite{FOR, Krasnov, OriRyan}.

\medskip 

This paper is a first step in the understanding and the analysis of the work of \cite{Freidel} in the hamiltonian context.
The hamiltonian
quantization is a very interesting quantization scheme because the notions of quantum states and quantum operators are very well 
understood in that framework. Therefore, one can compute explicit transition amplitudes whereas spin-foam models give a priori only
the partition function of the theory. The main problems with the hamiltonian quantization of gravity is that it breaks the covariance
of general relativity: there is an explicit choice of time slicing. Topology changings are not allowed because the topology of the
spacelike surface is a priori fixed once and for all. On the contrary, spin-foam models propose a covariant quantization 
of general relativity; this makes this approach so attractive. 
In this article, we reformulate the Loop Quantum Gravity (LQG) 
description of 3D gravity coupled to massive point particles where we make clear the crucial role
of quantum groups in the quantization. Therefore, we end up with a clear description of the $n$-particles states evolving in a quantum
background which is an essential step toward the hamiltonian description of a self-gravitating quantum scalar field theory. The last
point is presented in a companion paper \cite{Fockspace}.

\medskip

The paper is organized as follows. After the introduction, we start by recalling the techniques used in \cite{Noui1, Noui2} to perform
the hamiltonian quantization \`a la LQG of three dimensional gravity (pure or coupled to particles). In particular, we clarified the link
between the canonical quantization and the covariant spin-foam quantization. However, in this previous works, we never underline the role
of the quantum double $DSU(2)$ in our construction whereas it is clear that it should play a central role 
\cite{Freidellouapre, KM, Krasnov}. We fill this gap within the present article.
To that aim, we first define in the second section the notions of partial kinematical and physical Hilbert space. These spaces are
constructed from a particular  observer: this observer is a given particle which defines a reference frame where the momenta of the other 
particles and the energy of the global system are defined. This decription will appear very important above all in the construction of the
Fock space \cite{Fockspace}. In the third section, we show how the quantum double $DSU(2)$ is naturally present in the (partial) 
physical Hilbert space: (partial) physical states are defined as vectors of a tensor product of simple representations of $DSU(2)$
and the (partial) physical scalar product is nothing but an intertwining coefficients between simple representations of $DSU(2)$.
In this section, we also discuss some aspects concerning the observables: definition, computation of expectation values etc...
We present the construction first in the case of the sphere then we generalize to the case of any riemann surface.
We conclude with some perspectives and we postpone the construction of the Fock space to thr companion paper \cite{Fockspace}.

\medskip

In this paper, we will work with dimensionless quantities, in particular for the masses of the particles. But, the natural mass unit
is the Planck mass $m_P$ defined from the Newton constant $G$ by $m_P=1/(4\pi G)$. Then, any mass value has to be understood in terms
of the Plank unit. 

\subsection*{2. Space of Physical States}
Loop quantum gravity is presented as a two steps quantization of general relativity: the construction of the kinematical Hilbert
space and the construction of the physical Hilbert space. The first point is completely under control and gave rise to what is commonly
called quantum geometry (for review see \cite{Ashtekar}, \cite{Perez}, \cite{Rovelli}  or \cite{Thiemann}): 
the kinematical states (or states of quantum geometry) are described in term of spin-networks
and their set is endowed with a Hilbert structure defined by the Ashtekar-Lewandowski measure. But the construction of the physical 
Hilbert space is one the biggest difficulties and one of the most interesting problem of LQG. Different ways have been 
explored to attack the problem: construction of the extractor $P$ by using a regularization of the hamiltonian constraint 
\cite{RovGaull,Hamiltonianconstraint},
the master constraint program \cite{Masterconstraint}, spin-foam models \cite{Oriti,Perez2} etc... 
Very recently, Thiemann \cite{Existenz} and  Han $\&$ Ma \cite{Chinese} proved the existence of the 
physical Hilbert space in the context of the master program using spectral techniques. 
Nevertheless, no consistent solution have been found so far!

In a couple of papers \cite{Noui1, Noui2}, we have successfully studied 3D general relativity (pure gravity and gravity coupled to
particles) as a toy-model for LQG. In particular, we have constructed the extractor $P$ and have shown that the physical
scalar product between states has a spin-foam representation. Thus, we made the explicit relation between LQG and spin-foam models. 

This section is devoted to recall the construction in the pure gravity case
and in the case where gravity is coupled to massive spinless point particles. Then, we propose an alternative description 
for the physical Hilbert space from where the quantum double structure emerges naturally.

\subsubsection*{2.1. Pure Gravity: review and notations}
Let $\cal M$ be a three dimensional manifold. 
In the first order formalism, the degrees of freedom of the gravitational field in $\cal M$ are encoded in a triad 
$e(x)=e_\mu^i(x)J_i dx^\mu$ and in a spin-connection $\omega(x)=\omega_\mu^i (x)J_i dx^\mu$ where $J_1,J_2,J_3$ are the generators of the 
Lie algebra ${\mathfrak g}
={\mathfrak su}(2)$ satisfying the Lie-algebra relation $[J_i,J_k]=\varepsilon_{ij}{}^l J_l$. 
Indices are lowered and raised by the kronecker symbols $\delta_{ij}=\delta^{ij}$.
The dynamics is governed by the $BF$-action for the group $G=SU(2)$.

When the manifold has the topology ${\cal M}=\Sigma \times I$, $\Sigma$ being a Riemann surface of genus $g$ and $I$ a compact part of
the real line $\mathbb R$, then gravity admits a hamiltonian formulation. The non-reduced phase space is (schematically) defined by:
\begin{eqnarray}
{\cal E} \; \equiv \; \{(A,E) \; \vert \; A\; \text{a G connection on $\Sigma$} \;\; \text{and} \; E \; 
\text{an electric field on $\Sigma$}\}\;.
\end{eqnarray}
The variables $A$ and $E$ form a canonical variables pair and are respectively defined by the pullbacks of the spin-connection
and of the triad on the surface $\Sigma$.
The theory admits first class constraints that generate infinitesimal symmetries in the non-reduced phase space $\cal E$. 
An element $X \in {\cal E}$ is called $\cal G$-invariant when it Poisson-commutes with the constraints.
The symmetry group is given by ${\cal G}={\cal C}^\infty(\Sigma;ISU(2))$ where $ISU(2)$ is the (universal cover of the)
group of Euclidean transformations. It is a semi-direct group:
\begin{eqnarray}
ISU(2) \; = \; SU(2) \ltimes \mathbb R^3 \;\;\; \text{where $G$ acts on $\mathbb R^3$ by 
the vectorial representation}.
\end{eqnarray}
Note that the symmetry group $\cal G$ contains (on-shell) the diffeomorphisms group on $\Sigma$ but is not the diffeomorphism group.
Important discrepencies between gravity and $BF$ theory appears because of this fact \cite{Matschull}.  

The physical phase space $\cal P$ is obtained by applying an infinite dimensional version of the Dirac reduction to the space $\cal E$
and schematically reads:
\begin{eqnarray}
{\cal P} \; \equiv \; \{X \in {\cal E} \; \vert \; X  \; \text{is $\cal G$-invariant}\}/{\cal G}\;.
\end{eqnarray}
The space $\cal P$ is a finite dimensional symplectic manifold and it is isomorphic to the moduli space of flat $ISU(2)$-connections on the
surface $\Sigma$. Functions on $\cal P$ are called observables and, by definition, are invariant under the (induced) action of the 
symmetry group $\cal G$. 

\medskip

Different strategies have been developed the last fifteen years to quantize the theory (see \cite{Carlip} and references therein).
Most of them (combinatorial
quantization \cite{Alekseev, Buffenoir, BNR}, fonctional quantization \cite{Witten}, Nelson-Regge quantization \cite{Nelson} etc...) 
are proper to 3D gravity and cannot
be generalized to higher dimensions. What makes the loop and the spin-foam quantizations very attractive is precisely that
these schemes can, in principle, be applied in 4D. Even if neither LQG nor spin-foam models provide so far a complete
and consistent quantization  of 4D gravity. 

The loop quantization program consists in first quantizing the non-physical phase space $\cal E$, then promoting the first class
constraints into quantum operators and then finding the kernel of these quantum operators. In 4D gravity, all the constraints but one
can be solved in the quantum theory. The remaining and problematic constraint is known as the hamiltonian constraint.
In the context of 3D pure gravity, this program can be completely achieved and one can find solutions of all the constraints \cite{Noui1}.
We proceed as follows.

\newpage

\begin{enumerate}
\item {\bf Space of cylindrical functions on $\Sigma$: $\text{Cyl}(\Sigma,G)$.} 

We start by choosing the connection $A$ to be the configuration variable (choice of polarization) and we denote $\cal A$ the space 
of $G$-connections on $\Sigma$. We introduce the space of discrete connections associated to a graph $\gamma$ (with
$E$ edges and $V$ vertices), denoted
${\cal A}_\gamma$ by, rougthly speaking, replacing local connections $A$ with holonomies $U_e$ of the connection 
along the edges $e$ of the graph.
A cylindrical function on the graph $\gamma$ is an element $\psi \in Fun({\cal A}_\gamma) \subset Fun(\cal A)$ 
such that there exists a function $f \in Fun(G^{\otimes E})$ and $\psi(A)=f(\otimes_{e=1}^E U_e)$. The space of cylindrical functions
is denoted $\text{Cyl}(\gamma;G)$ and is naturally endowed with a measure $d\mu_\gamma = \otimes_{e=1}^E d\mu_e$ where $d\mu_e$ is the
normalized $G$ Haar measure associated to the edge $e$. 
Finally, we define the space of cylindrical functions on the surface $\Sigma$ as the following union:
\begin{eqnarray}
\text{Cyl}(\Sigma;G) \; = \; \bigcup_\gamma \text{Cyl}(\gamma;G)\;.
\end{eqnarray}
This space inherits a natural measure, known as the Ashtekar-Lewandowski measure, and the completion of $\text{Cyl}(\Sigma;G)$ with
respect to this measure is the Hilbert space of non-physical states, denoted ${\cal H}(\Sigma;G)$. The action of the symmetry group
$\cal G$ on the connections induces a (co-) action on $\text{Cyl}(\Sigma;G)$ and on ${\cal H}(\Sigma;G)$ and we will use the same notation
$\cal G$ to denote the symmetry group acting on $\text{Cyl}(\Sigma;G)$.
Physical states are those functions which are left invariant under this symmetry. 

\item {\bf Invariance under the symmetry group $G$.}

The action of the symmetry sub-group $C^\infty(\Sigma;G)$ reduces to an action on the vertices of any element of ${\cal A}_\gamma$
for any graph $\gamma$. Invariant states under this action are called kinematical states. The set of kinematical states is a 
vector subspace of $\text{Cyl}(\Sigma;G)$ and, due to left and right invariance of the $G$ Haar measure, inherits the pre-Hilbert
structure of $\text{Cyl}(\Sigma;G)$. After completion, we obtain the kinematical Hilbert space, denoted ${\cal H}_{kin}(\Sigma;G)$.
Spin-network states provide an orthonormal basis of ${\cal H}_{kin}(\Sigma;G)$.

\item {\bf Invariance under the translations $\mathbb R^3$.}

Physical states are a priori kinematical states that are invariant under the action of the residual symmetry subgroup. 
However, there is no physical states in the kinematical Hilbert space. This is a well-known fact that is a consequence of the 
non-compactness of the sub-group $\mathbb R^3$. Physical states are in fact ``distributional'', in the sense that they are elements of 
$\text{Cyl}(\Sigma;G)^*$, the topological dual of $\text{Cyl}(\Sigma;G)$. Finding physical states is equivalent to finding a 
physical extractor (abusively called a projector), i.e. an operator:
\begin{eqnarray}
P \; : \; \text{Cyl}(\Sigma;G) \; \longrightarrow \; \text{Cyl}(\Sigma;G)^*
\end{eqnarray}
that satisfies the property $P(\psi_1)(\psi_2)=P(\psi_1)(\xi \cdot \psi_2)$ for any cylindrical functions $\psi_1$ and $\psi_2$
and any element $\xi \in \cal G$. We have denoted by $\cdot$ the action of $\cal G$ on $\text{Cyl}(\Sigma;G)$. Thus 
the physical Hilbert space ${\cal H}_{phys}(\Sigma;G)$ is defined as the image of $\text{Cyl}(\Sigma;G)$ by $P$. 
Given two physical
states $\phi_1$ and $\phi_2$, there exists $\psi_1, \psi_2 \in \text{Cyl}(\Sigma;G)$ such that $\phi_1=P(\psi_1)$ and 
$\phi_2=P(\psi_2)$, and the physical scalar product is defined by:
\begin{eqnarray}
<\phi_1,\phi_2>_{phys} \; \equiv \; <\psi_1,\psi_2>_{phys} \; \equiv \; P(\psi_1)(\psi_2) \;.
\end{eqnarray}
The physical scalar product does not depend on the choice of the functions $\psi_1$ and $\psi_2$. Generally, we 
omit to mention the elements $\phi_1$ and $\phi_2$, and we write the physical scalar product between cylindrical functions.
The operator $P$ ``extracts'' the physical component of any cylindrical function.
\end{enumerate}

\begin{figure}[h]
\psfrag{psi=}{$\psi(j_1,\cdots,j_6)=$}
\psfrag{sca}{$<\psi\vert\emptyset>_{phys}=$}
\psfrag{j1}{$j_1$}
\psfrag{j2}{$j_2$}
\psfrag{j3}{$j_3$}
\psfrag{j4}{$j_4$}
\psfrag{j5}{$j_5$}
\psfrag{j6}{$j_6$}
\centering
\includegraphics[scale=0.7]{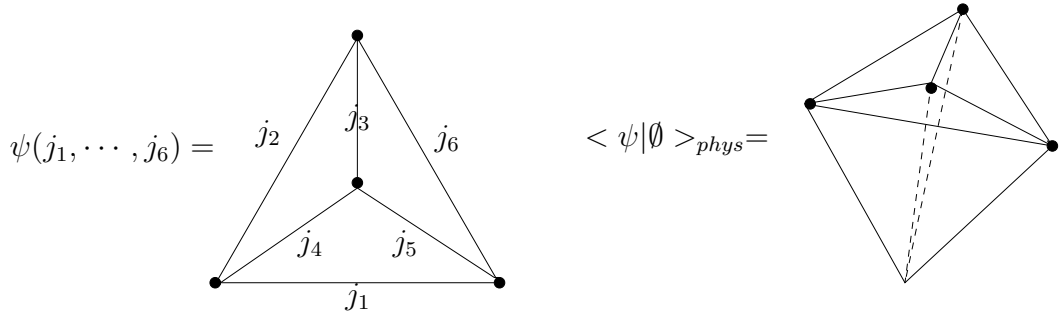}
\caption{\it The picture on the left is a spin-network state whose edges are labelled with irreps of $SU(2)$ and vertices with normalized
$SU(2)$ intertwiners. On the right side, we have drawn the spin-foam picture illustrating the transition amplitude between the no-state
and the state on the left. The amplitude associated to this spin-foam is a $(6j)$ symbol.}
\label{startriangle}
\end{figure}

The explicit construction of $P$ has been done in \cite{Noui1}. In fact, the extractor has been constructed by imposing locally
the flatness condition $F(A)=0$ after a choice of regularization. The physical scalar product so obtained is well-defined (i.e.
convergent), is positive and satisfies hermiticity condition. Moreover, the relation to spin-foam model has been established and is
breefly recalled in the following. Given two spin-network states $\psi_1$ and $\psi_2$ respectively associated to the colored graphs 
$\gamma_1, \gamma_2 \subset \Sigma$, one associates a cellular decomposition $\Delta$ of the three-dimensional manifold 
$\Sigma \times I$ which interpolates the graphs $\gamma_1$ and $\gamma_2$ at the boundaries. Moreover, $\Delta$ is colored in the
usual way: faces are associated with irreps of $SU(2)$, edges with intertwiners and the colors are compatible with the boundaries
colors. At this point, one introduces the dual graph $\Gamma$ whose edges are now colored with $SU(2)$ irreps.
The spin-foam amplitude associated to the graph $\Delta$ (or in an equivalent way to the dual graph $\Gamma$)
gives the physical scalar product between the two spin-network states if and
only if $\Delta$ is free of bubles, i.e. $\Delta$ has a tree structure. The last condition makes the amplitude convergent and 
appears naturally in the hamiltonian framework as a consequence of the regularization of the constraint $F(A)=0$. Therefore,
in the case where $\Gamma$ is a triangulation of $\Sigma \times I$ we have:
\begin{eqnarray}\label{PRmodel}
<\psi_1,\psi_2>_{phys} \; = \; \sum_{j_e} \prod_{e\in \Gamma} W_e(\{j_e\}) \; \prod_{t\in \Gamma} W_t(\{j_e\})
\end{eqnarray}
where the sum runs over $SU(2)$ irreps coloring the edges $e$ of $\Gamma$, the weight $W_e$ is generically given by $d_{j_e}=2j_e+1$ 
and is $d_{j_e} \delta_{j_e,j_\ell}$ if the edge $e$ is dual to a face whose boundary contains an edge of $\gamma_1$ or $\gamma_2$
colored by $j_\ell$; $W_t=(6j)$ is the normalized symmetric $(6j)$ symbols defined by the six representations coloring the edges of 
the tetrahedron $t$.
This property is illustrated by the following example:
\begin{eqnarray}
<\emptyset \; \vert \; \psi(j_1,\cdots,j_6)>_{phys} \; = \; N(j_1,j_2,j_4) N(j_2,j_3,j_6) N(j_1,j_5,j_3)
\left\{ \begin{array}{ccc} j_1 & j_2 & j_6 \\ j_4 & j_5 & j_3 \end{array} \right\}\;.
\end{eqnarray}
where $N(j_1,j_2,j_3)$ is a the norm of three-valent intertwiners (evaluation of $\Theta$ spin-network) taken to be 1.
The state $\psi(j_1,\cdots,j_6)$ is the spin-network state defined on the sphere and 
represented in the picture (\ref{startriangle}).

Let us finish with a remark.
We consider two cylindrical functions $\psi_1$ and $\psi_2$ which differ only by the fact that their associated graphes
are different but related by a diffeomorphism. Therefore, the state $P(\psi_1) - P(\psi_2)$ is a null-vector in the physical Hilbert space
and we can identify the spaces $P(\text{Cyl}(\gamma;G))$ and $P(\text{Cyl}(\gamma';G))$. This is a consequence of the invariance of
the theory under the diffeomorphisms group. Furthermore, the identification of the spaces still holds when the graphs $\gamma$ and 
$\gamma'$ are homotopic and not necessary related by a diffeormorphism. Thus, we identify states that are not related by a diffeomorphism
when we impose the flatness condition. We have to be aware of this fact which is a consequence of the fact that we are working with
first order gravity and then we include degenerate metrics in the model. How to avoid degenerate metrics in the quantum theory is still
an open question.

\subsubsection*{2.2. Coupling to a finite number of particles}
In order to couple particles to the gravitational field, we now consider a surface $\Sigma$ with boundary: the genus of $\Sigma$ is 
still denoted
$g$ and we assume the boundary is a disjoint union of $n$ boundaries whose topology is a circle, i.e. $\partial \Sigma=\bigcup_{i=0}^{n-1}
 {\cal B}_i$.
Each boundary ${\cal B}_i$ is associated to one particle ${\cal P}_i$ whose degrees of freedom are encoded in an element $X_i=(\Lambda_i,q_i) 
\in ISU(2)$:
$q_i \in \mathbb R^3$ is the position of the particle and $\Lambda_i \in G$ is related to the momemtum by $\vec{p}_i = m_i \Lambda_i 
\vec{n}$ where $\vec{n} \in S^2$. 
The mass $m_i$ of each particle ${\cal P}_i$ is fixed by appropriate boundary conditions of the gravitational field on ${\cal B}_i$
\cite{BN,DeSousa}. 
The action of the coupled system is obtained via a minimal coupling and takes the form:
\begin{eqnarray}\label{couplingaction}
S \; = \; S_{BF}[e,\omega] \; + \; \sum_{i=0}^{n-1} S_{P_i}[q_i,\Lambda_i] \; + \; S_C[e,\omega;(q_i,\Lambda_i)_i] \;.
\end{eqnarray}
The explicit expression of the action is given in \cite{Noui2}: $S_{BF}$ is the usual $SU(2)$ $BF$-action, $S_P$ is the (first order)
action of free relativistic particles and $S_C$ represents the minimal coupling term. Note that $S_C$ enforces the appropriate 
boundary conditions which fixes the masses of the particles.

The loop quantization of the action (\ref{couplingaction}) is conceptually similar to the pure gravity case
and has been performed in \cite{Noui2}. Let us recall the main results.
\begin{enumerate}
\item {\bf Particles-cylindrical functions.}

Particles-cylindrical functions are a direct generalization of the usual notion of cylindrical functions. A particles-cylindrical function $\psi$ is
now a function of the connection $A$ and of the momenta $\Lambda_i$. Such a function is
defined by a graph $\gamma \subset \Sigma$ which can admit open ends at the boundaries and by a function $f \in Fun(G^{\otimes (E+O)})$ where $E$ is 
the number of edges and $O$ the number of open ends such that:
\begin{eqnarray}
\psi(A,\Lambda_i) \; = \; f(\otimes_{e=1}^E U_e \otimes_{o=1}^O \Lambda_o) \;.
\end{eqnarray}
Remind that $U_e$ is the holonomy of the connection along the edge $e$ and $\Lambda_o=\Lambda_i$ if the open end $o \in {\cal B}_i$. 
Therefore, the vector space of particles-cylindrical functions $\text{Cyl}(\Sigma,{\cal P};G)$ is given by:
\begin{eqnarray}
\text{Cyl}(\Sigma,{\cal P};G) \; \equiv \; \text{Cyl}(\Sigma;G) \otimes Fun(SU(2))^{\otimes n} \;.
\end{eqnarray}
${\cal P}$ labels the set of the particles. This space is naturally
endowed with a measure (constructed from the AL measure and the $SU(2)$ normalized Haar measure) which makes it a pre-Hilbert space. Its
completion is denoted ${\cal H}(\Sigma,{\cal P};G)$. 
\item {\bf Symmetries of the system.}

The system admits first class constraints that generates infinitesimal symmetries. The symmetry group of the Hamiltonian theory is a cartesian product 
${\cal G}\times {\cal C}$. Any element of ${\cal G}$ is an element of $C^\infty(\Sigma,ISU(2))$ which is constant at each boundary ${\cal B}_i$. As in the pure gravity case, we distinguish the ``compact'' subgroup $C^\infty(\Sigma,SU(2))$ from the non-compact one 
$C^\infty(\Sigma,\mathbb R^3)$. 
The compact part generates usual gauge transformations which read on the variables $A$ and $\Lambda_i$:
\begin{eqnarray}\label{calG}
\forall \; g \in {\cal G} \;\; \text{s.t.} \; g\vert_{{\cal B}_i} = g(i) \;\;\; (A,\Lambda_i) \; \mapsto \; (g^{-1} A g + g^{-1}dg,\; 
g(i)^{-1}\Lambda_i). 
\end{eqnarray}
The non-compact part is related to the diffeomorphisms group of the manifold $\Sigma \times I$ and acts non-trivially
on the momenta canonically conjugated to the connection $A$. 

The group ${\cal C}=Cst(\partial \Sigma,\Gamma)$ is the set of functions on $\partial \Sigma$ which are constant on each components ${\cal B}_i$
and take value in the Cartan torus $\Gamma$ of $ISU(2)$ (Note that $\Gamma$ is the two-dimensional group $\mathbb R \times S^1$).
Therefore ${\cal C} \simeq \Gamma^{\times n}$. It acts trivially on the connection $A$ and
by right multiplication on the momenta as shown in the following:
\begin{eqnarray}\label{calC}
\forall \; (\lambda(i),\tau(i))_{i\in[0,n-1]} \in {\cal C} \;\; \text{s.t.} \; \lambda(i) \in SU(2) \;, \; \tau(i) \in \mathbb R^3 \;\;\;\;\;\;\;\;\;
\Lambda_i \; \mapsto \; \Lambda_i h(i) \;.
\end{eqnarray}
One can show that $\mathbb R \subset \Gamma$ is the subgroup of
time reparatrizations of the worldline of each particle whereas $S^1 \subset \Gamma$ is the subgroup of internal 
frame rotations which preserves the
momentum of each particle. It is then natural that only the $SU(2)$ Cartan subgroup has a non-trivial action on the momenta.

The action of $\cal G$ induces a co-action on $\text{Cyl}(\Sigma,{\cal P};G)$ and particles-physical states are
the elements of $\text{Cyl}(\Sigma,{\cal P};G)$ which are co-invariant under $\cal G$. 
\item {\bf Imposing the constraints and the particles-physical Hilbert space.}

Invariance under reparametrization is trivial because we work in the momentum representation
and invariance under internal frame rotations imposes that states are functions on the sphere $S^2=SU(2)/U(1)$ instead of a function
on the whole group $SU(2)$ (see \cite{Noui2} for more details). Therefore, the set of particles-cylindrical functions satisfying these
invariances reads:
\begin{eqnarray}
\text{Cyl}_{{\cal C}}(\Sigma,{\cal P};G) \; \equiv \; \text{Cyl}(\Sigma,{\cal P};G) / {\cal C}
\; = \; \text{Cyl}(\Sigma;G) \otimes Fun(S^2)^{\otimes n}\;.
\end{eqnarray}
Due to the left and right invariance of the $G$ Haar measure, the space $\text{Cyl}_{\cal C}(\Sigma,{\cal P};G)$ inherits the measure of
$\text{Cyl}(\Sigma,{\cal P};G)$.
To find particles-physical states, we have to impose the invariance under $\cal G$. As in the pure gravity case, this is a two-steps
procedure. First, we impose the invariance under the sub-group of gauge transformations: this is immediate to do and we obtain the 
particles-kinematical Hilbert space ${\cal H}_{kin}(\Sigma,{\cal P};G)$. Particles-spin-network states provide an orthonormal basis of 
${\cal H}_{kin}(\Sigma,{\cal P};G)$. Note that particles-spin-network states are associated to graphs with (eventually)
open ends at the location of
the particles.

Imposing the remaining constraint (which generates the non-compact symmetry group)
is done in the same way as in the pure gravity case. This means finding an extractor $P$:
\begin{eqnarray}\label{Pparticles}
P \; : \; \text{Cyl}_{\cal C}(\Sigma,P;G) \; \longrightarrow \; \text{Cyl}_{\cal C}(\Sigma,P;G)^* 
\end{eqnarray}
such that $P(\psi_1)(\psi_2)=P(\psi_1)(\xi \cdot \psi_2)$ for any particles-cylindrical functions $\psi_1$ and $\psi_2$ and any
$\xi \in \cal G$. The image of $\text{Cyl}_{\cal C}(\Sigma,{\cal P};G)$ by the extractor is a vector space. We endow this space with a
(Ashtekar-Lewandowski) measure and
after completion we end up with the particles-physical Hilbert space ${\cal H}_{phys}(\Sigma,{\cal P};G)$.
\end{enumerate}
The explicit construction of $P$ has been done in \cite{Noui2}. The relation to spin-foam models has been unraveled: the 
particles-physical scalar product reproduces spin-foam amplitudes of \cite{Freidellouapre} as shown in the following section.
It would be very interesting to ask the question of the unicity of the extractor $P$ in the context
of three dimensional gravity because it seems that the same kinematical space is the starting point of the construction of 
different physical states depending whether we are dealing with pure gravity, gravity coupled to particles or in the presence of a
cosmological constant...

\subsubsection*{2.3. The physical states}
As presented before, any physical state can be viewed as an equivalent class of a particles-cylindrical function
(two states are equivalent if their difference has a non-zero physical norm) or equivalently as the image of a  
particles-cylindrical function by the extractor $P$ defined above (\ref{Pparticles}).
A particles-physical state $\psi$ is said to be explicitely dependent of the particles degrees of freedom if any of its representative
is a particles-cylindrical function defined on a graph which has at least one open vertex (at a boundary). Otherwise, the state does not
depend explicitely on the particles-degrees of freedom.

States which do not depend explicitely on the particles degrees of freedom caracterize the three dimensional space-time geometry
with conical singularities but do not contain any information (a part from the masses) concerning the particles. The physical scalar
product between two such (spin-network) states $\psi_1$ and $\psi_2$ has a regularized well-defined spin-foam representation of the 
type \cite{Freidellouapre}. Indeed, one fixes a tridimensional graph $\tilde{\Gamma} \subset \Sigma \times I$ 
with no-buble interpolating the flat graphes 
$\gamma_{1}$ and $\gamma_{2}$ associated to the cylindrical functions $\psi_1$ and $\psi_2$ and the dual graph $\Gamma$, and we still have
the identity (\ref{PRmodel}).
In that case, edges $e$ of $\Gamma$ are still colored with a representation $j_e$. The weight associated to each tetrahedron is
still given by $W_t(\{j_e\})=(6j)_t$ but the weight $W_e$ associated to each edge is slightly modified:
$W_e  =  \chi_{j_e}(m_i)$  where $\chi_{j}(m)\equiv \sin(2j+1)m/\sin m$
if the boundary associated to the particle of mass $m_i$ crosses the dual face $f$ of the edge $e$;
$W_e= 2j_\ell +1$ where $j_\ell$ is the representation coloring one edge $\ell$ of the graphes 
$\gamma_{\varphi}$ or $\gamma_\psi$ if the boundary of the dual face of $e$ contains the edge $\ell$;
 $W_e=2j_e+1$ otherwise.
Note that, in the construction of \cite{Freidellouapre}, one has to fix a maximal tree $T$ of $\Gamma$ and to impose the condition
$W_e=\delta_{j_e,0}$ to each edge $e$ of $T$ in order to make the sum (\ref{PRmodel}) convergent. In our construction, we do not have
such a condition because the graph $\tilde{\Gamma}$ is chosen to be free of bubbles. The two conditions are in fact equivalent.

\begin{figure}[h]
\psfrag{I}{$j_1$}
\psfrag{J}{$j_2$}
\psfrag{K}{$j_3$}
\psfrag{L}{$j_4$}
\psfrag{M}{$j_5$}
\psfrag{N}{$j_6$}
\psfrag{m}{$m$}
\centering
\includegraphics[scale=0.7]{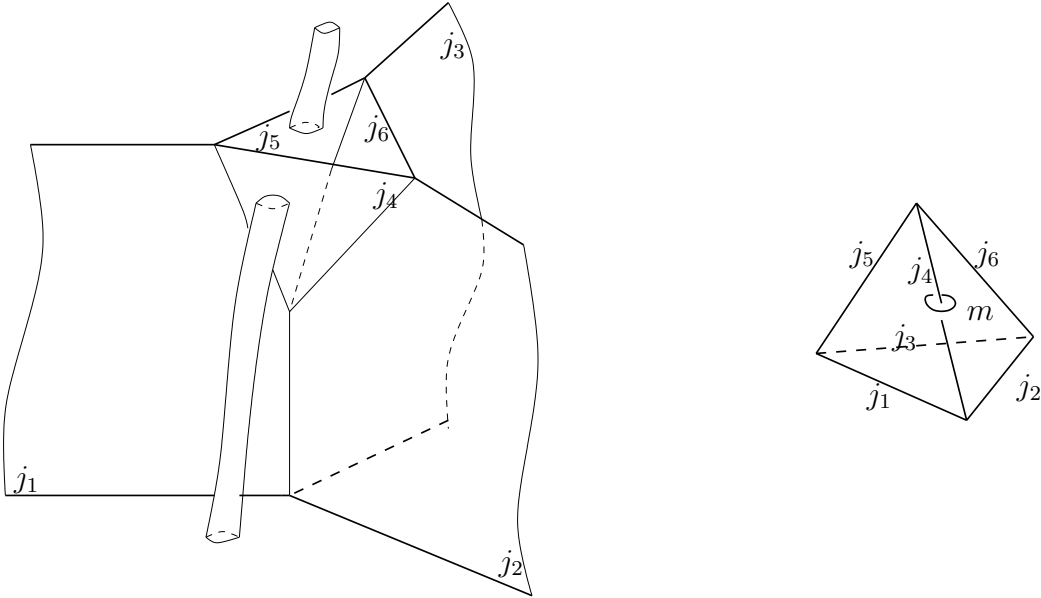}
\caption{\it On the l.h.s. chunk of spin-foam amplitude between two spin-network states represented in thick lines: the boundary
crosses the face labelled with the irrep $j_4$. On the r.h.s. the dual picture: the circle around the edge $j_4$ reminds the presence
of the boundary.}
\label{example}
\end{figure}

The relation (\ref{PRmodel}) is illustrated in the example of the picture (fig. \ref{example}).

The spin-foam amplitude associated to the spin-foam on the l.h.s. is given by:
\begin{eqnarray}
\frac{\chi_{j_4}(m)}{d_{j_4}}\; \left\{ \begin{array}{ccc} j_1 & j_2 & j_6 \\ j_4 & j_5 & j_3 \end{array} \right\}\;.
\end{eqnarray}

Reciprocally, the amplitude of any spin-foam of the type \cite{Freidellouapre} can be interpreted as the physical scalar 
product between two states which do not depend explicitely on the particles degrees of freedom.

\begin{figure}[h]
\psfrag{I}{$I$}
\psfrag{J}{$J$}
\psfrag{K}{$K$}
\centering
\includegraphics[scale=0.7]{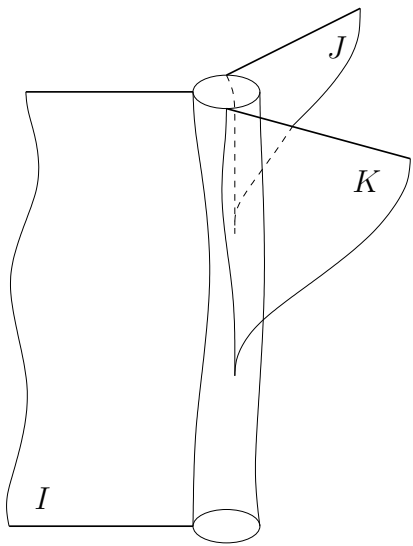}
\caption{\it Example of transition between spin-network states involving explicitely boundaries degrees of freedom. 
New faces can emerge from the boundaries.}
\label{example2}
\end{figure}

Of particular interest is the physical scalar product between states which depend explicitely on the particles degrees of freedom. 
This kind of physical scalar product does not admit a spin-foam representation of the previous type \cite{Freidellouapre}.
In that case, the cellular decomposition of the manifold $\Sigma \times I$, defining the eventual spin-foam model, involves faces
whose edges belong to the boundaries (see figure \ref{example2} as an example) whereas the faces where crossed by the boundaries in the 
previous case. 

It would be very interesting to generalize the spin-foam model developed in \cite{Freidellouapre} in order to include transitions
of the type (figure \ref{example2}). We let this issue for future investigations. In the sequel, we concentrate on the physical
scalar product between particles-physical states and we do not ask for the moment the question of its spin-foam representation.
Then, we want to chose a basis of physical states and we are going to compute the physical scalar product between the elements
of the basis. In other words we are going to compute explicitely the matrix elements of the extractor $P$ whose definition
and properties were briefly recalled above (\ref{Pparticles}).

\medskip

\begin{figure}[h]
\psfrag{x}{$x$}
\psfrag{e1}{$e_1$}
\psfrag{e2}{$e_2$}
\psfrag{a1}{$a_1$}
\psfrag{a2}{$a_2$}
\psfrag{b1}{$b_1$}
\psfrag{b2}{$b_2$}
\centering
\includegraphics[scale=1.0]{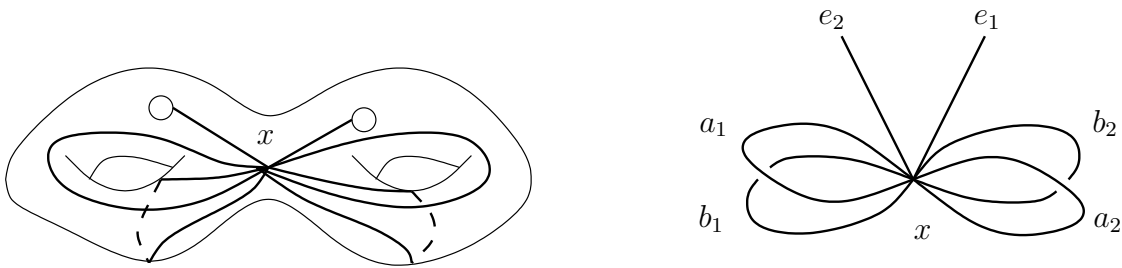}
\caption{\it A minimal graph on a genus 2 surface with two particles. On the r.h.s., the ``flat projection'' of the minimal graph
with the labellings of each edge or loop.}
\label{minimalgraph}
\end{figure}

For that purpose, we start by chosing a minimal graph $\gamma$ on the Riemann surface with boundaries $\Sigma$: $g$ and $n$
denote respectively the genus of $\Sigma$ and the number of connected components of $\partial \Sigma$. A minimal graph 
(figure \ref{minimalgraph}) consists in
a choice of a base point $x \in \Sigma$, a choice of $n$ oriented edges linking $x$ and a point on each boundary ${\cal B}_i$ 
(let $E$ be the set of these edges) and a choice of $2g$ oriented non-contractible loops around each handles of the surface
(let $H=A\cup B$ be the set of these loops that we separate as usual into two sets $A$ and $B$). 

The space of discrete connections on the minimal graph is denoted ${\cal A}_\gamma$: any discrete connection is a family 
$A=(U_a,U_b;U_e,\Lambda_i)_{a,b,e,i} \in G^{2(g+n)}$ where $e \in [0,n-1]$ labels the elements of $E$, $a,b \in [1,g]$ 
the elements of $H=A\cup B$ and $i \in [1,n]$ the particles. The infinite dimensional symmetry group 
$\cal G\times {\cal C}$ (\ref{calG},\ref{calC}) acting on the space of connections reduces to a finite dimensional group 
when acting on ${\cal A}_\gamma$. The compact part of this group is trivially isomorphic to $G \times (G\times S^1)^{\times n}$. 
Its action on ${\cal A}_\gamma$ is given by:
\begin{eqnarray}
(G\times (G\times S^1)^{\times n}) \; \times \; {\cal A}_\gamma & \longrightarrow & {\cal A}_\gamma  \\
(y,(g(i),\lambda(i))_i) \; \times \; (U_a,U_b;U_e,\Lambda_i) & \longmapsto & (y^{-1}U_ay,y^{-1}U_by;y^{-1}U_e g(i),g(i)^{-1}
\Lambda_i \lambda(i))  \;.\nonumber
\end{eqnarray}
The non-compact subgroup acts trivially on the variables $A$ and $\Lambda_i$ and has a non-trivial action on the variables
canonically conjugated. Imposing invariance under the non-compact subgroup is done by finding the appropriate extractor $P$.

The functions on discrete connections form the set $\text{Cyl}(\gamma;G)$ of cylindrical functions on the minimal graph $\gamma$.
Recall that $\psi \in \text{Cyl}(\gamma;G)$ if there exists a function $f \in Fun(G^{2(g+n)})$ such that given $A \in {\cal A}_\gamma$
we have:
\begin{eqnarray}
\psi(A) \; = \; f(\bigotimes_{a,b=1}^g U_a \otimes U_b \otimes \bigotimes_{e,i=0}^{n-1} U_e \otimes \Lambda_i)\;.
\end{eqnarray}
Due to the compactness of the gauge group $G=SU(2)$, it is clear that $\text{Cyl}(\gamma;G)$ is isomorphic to the space 
$Fun(G)^{2(g+n)}$. Then, the space of cylindrical functions is naturally endowed with a pre-Hilbert structure defined by 
the Haar measure on the group $G$ and its completion, denoted ${\cal H}(\gamma;G)$, is called the space of non-physical states.
This space is essential in the constructions of the kinematical Hilbert space and the physical Hilbert space.
Obviously, these constructions do not depend on the  choice of the base point $x$ on the surface: indeed, the base point has no physical 
meanning and it is natural that ``physical'' quantities do not depend on it. Technically, it is the adjoint action of the gauge group
$G$ which makes the physical Hilbert space independant of the base point.
Even the kinematical Hilbert space does not depend on that choice.

Therefore, one can choose $x$ to be a point of a given boundary, let us say ${\cal B}_0$. Yet, 
there exists now two types of symmetries acting on the variables defined on ${\cal B}_0$: the action of $G$ and the action of $S^1$. 
The former is the usual action of the gauge group $SU(2)$ whereas $S^1$ is the group of internal frame symmetries (associated to the 
particle ${\cal P}_0$). To construct kinematical states, one has to consider both symmetries and find invariant functions under these
two symmetries. However, it is interesting to forget the $S^1$ invariance. This is equivalent to choose a particular internal frame
for the particle ${\cal P}_0$. Thus, one can capture interesting informations concerning the particles: in particular one can define
the momenta of each particles in the internal frame of the particle ${\cal P}_0$. This is obviously impossible if we do not choose
a particular frame. Thus, one can interpret the particle ${\cal P}_0$ as an observer sitting somewhere in the surface $\Sigma$.
This observer is measuring the physical caracteristics of each particles and also the properties of the spacelike geometries in its 
reference frame. $G$-invariant states defined in this reference frame are not kinematical states in the sense of LQG; they are called
in the sequel partial kinematical states. The set of partial kinematical states is naturally endowed with a pre-Hilbert structure.
Its completion is called the partial kinematical Hilbert space and is denoted ${\cal H}_{Pkin}$.
Note that one can extend the space ${\cal H}_{Pkin}$ to distributions in the sense that one can choose the element $\psi$ defining
the spin-network to be a distribution rather than a function. In that case, one can define pure momenta states for instance as ``delta''
functions. As the manifold in consideration ($G$ or $S^2$) are compact, we will identify the space of function with the space of 
distributions on this manifold. We keep this remark in mind for the following.

\begin{proposition}[Partial kinematical Hilbert space]\label{Pkin}
The Hilbert space of partial kinematical states ${\cal H}_{Pkin}$ is isomorphic to the space $Fun(G^{2g}\otimes (S^2)^{n-1})$
endowed with the measure $d\mu_G^{\otimes 2g} \otimes d\mu_{S^2}^{\otimes (n-1)}$. Indeed,
a function $\psi \in {\cal H}(\gamma;G)$ is a partial kinematical states if and only if there exists a 
function $f \in Fun(G^{2g}\otimes (S^2)^{n-1})$ such that:
\begin{eqnarray}
\psi(A) \; = \; f(\bigotimes_{a,b} \Lambda_0^{-1} U_a \Lambda_0 \otimes \Lambda_0^{-1} U_b \Lambda_0 \bigotimes_{e=1}^{n-1} 
\Lambda_0^{-1} U_e \tilde{\Lambda}_{e})\;\;,
\end{eqnarray}
where $\tilde{\Lambda}_e = \int dh \; \Lambda_e h \in S^2$ and $dh$ is the measure on the Cartan torus of $G$. 
We have used the notation $g\tilde{\Lambda}=\int dh \; g\Lambda h$ for the action of the group element $g \in SU(2)$
on a point $\tilde{\Lambda} \in S^2$.
\end{proposition}

{\it Proof.}

First, let us impose the invariance under the action of $G$ on the boundary ${\cal B}_e$ associated to any particle ${\cal P}_e$
but the observer ${\cal P}_0$. The group $G$ acts non-trivially on the degrees of freedom $\Lambda_e$ and $U_e$ according to the 
following maps (\ref{calG}):
\begin{eqnarray}
\forall \; g \in G \;\;\; \Lambda_e \; \longmapsto \; g\Lambda_e \;\;\;\;
 \text{and} \;\;\;\;\; U_e \; \longmapsto \; U_e g^{-1} \;.
\end{eqnarray}
This action induces a co-action on ${\text{Cyl}}(\gamma;G)$ and it is obvious that any function 
$\psi \in {\text{Cyl}}(\gamma;G)$
is co-invariant if and only if it is a function of $U_e\Lambda_e$. This results holds for any particle $e \neq 0$. The action of $G$
at the boundary ${\cal P}_0$ is more involved because there are many edges ending and starting at ${\cal P}_0$. 
But the conclusion
is similar because any function $\psi$ which is $G$
invariant at ${\cal P}_0$ can be written as a sum of functions of the variables $\Lambda_0^{-1}U_e$, $\Lambda_0^{-1} U_a \Lambda_0$ and
$\Lambda_0^{-1} U_b \Lambda_0$. To proove this point, we decompose the function $\psi$ into (tensor product of) irreducible unitary 
finite dimensional representations (irreps) of $G$ using the Plancherel formula and we obtain:
\begin{eqnarray}\label{decomposition}
\psi(A) \; = \; \sum_{[I],[j],[k]} \stackrel{[I]}{f}{}\!\!_{[j]}^{[k]}(\Lambda_e)
\prod_{\ell} \stackrel{I_\ell}{\pi}{}\!\!^{j_\ell}_{k_\ell} (U_\ell) 
\times \stackrel{I_0}{\pi}{}\!\!^{j_0}_{k_0} (\Lambda_0) \;.
\end{eqnarray}
$[I] \in (\frac{1}{2}\mathbb N)^{2g+n}$ are $SU(2)$ irreps labelling the edges $\ell$ (for the links $e$ and the 
non-contractible loops $a,b$) and the reference particle ${\cal P}_0$; $[j]$ and $[k]$ are families of magnetic numbers.
We have denoted $\stackrel{I}{\pi}\!\!^j_k=<\stackrel{I}{e}{}\!\!^j\vert \stackrel{I}{\pi} \vert \stackrel{I}{e}_k>$ the
matrix elements of irreps of $SU(2)$, $\stackrel{I}{V}$ is the vector space associated to the spin $I$ representation
and $\stackrel{I}{e}_i$ (resp. $\stackrel{I}{e}{}\!\!^i$) the (resp. dual) basis of $\stackrel{I}{V}$ (resp. $\stackrel{I}{V}{}^*$).
The dependence in the variables $\Lambda_e$, $e \neq 0$, is contained in the definitions of the Fourier components 
$\stackrel{[I]}{f}{}\!\!_{[j]}^{[k]}(\Lambda_e)$. The decomposition (\ref{decomposition}) is nothing else but the spin-network
decomposition of the state.

The $G$ invariance at ${\cal P}_0$ implies that there exist a $G$ intertwiner $\iota$ defined by:
\begin{eqnarray}
\iota \in \text{Hom}(\stackrel{I_0}{V}\otimes \bigotimes_{\ell=1}^{2g+n-1} \stackrel{I_\ell}{V}; \bigotimes_{a,b=1}^g
\stackrel{I_a}{V} \otimes \stackrel{I_b}{V}) \otimes (\stackrel{I_0}{V} \otimes \bigotimes_{e=1}^{n-1}\stackrel{I_e}{V})^* 
\;\;\;\;\text{such that} \nonumber \\
\stackrel{[I]}{f}{}\!\!_{[j]}^{[k]} \; = \; <\bigotimes_{a,b=1}^g
\stackrel{I_a}{e}{}\!\!^{k_a} \otimes \stackrel{I_b}{e}{}\!\!^{k_b} \vert 
\iota(\stackrel{I_0}{e}{}\!\!^{k_0} \otimes \bigotimes_{e=1}^{n-1}\stackrel{I_e}{e}{}\!\!
^{k_e}) \vert \stackrel{I_0}{e}_{i_0} \otimes  \bigotimes_{\ell=1}^{2g+n-1} \stackrel{I_\ell}{e}_{j_\ell} >\;.
\end{eqnarray} 
In the language of LQG, we color the vertex ${\cal P}_0$ with the intertwiner $\iota$.
Due to the invariance of the intertwiner $\iota$, we have:
\begin{eqnarray}
\psi(A) \; = \; \sum_{[I],[j],[k]} \stackrel{[I]}{f}{}\!\!_{[j]}^{[k]}(\Lambda_e)\; \delta^{j_0}_{k_0} \;
\prod_{a} \stackrel{I_a}{\pi}{}\!\!^{j_a}_{k_a} (\Lambda_0^{-1} U_a \Lambda_0)\;
\prod_{b} \stackrel{I_b}{\pi}{}\!\!^{j_b}_{k_b} (\Lambda_0^{-1} U_b \Lambda_0)\;
\prod_{e} \stackrel{I_e}{\pi}{}\!\!^{j_e}_{k_e} (\Lambda_0^{-1} U_e)\;.
\end{eqnarray}
Then, we impose the $G$ invariance at the location of each particle and we obtain in the same way that:
\begin{eqnarray}
\psi(A) \; = \; \sum_{[I],[j],[k]} \stackrel{[I]}{f}{}\!\!_{[j]}^{[k]}(1)\; \delta^{j_0}_{k_0} \;
\prod_{a} \stackrel{I_a}{\pi}{}\!\!^{j_a}_{k_a} (\Lambda_0^{-1} U_a \Lambda_0)\;
\prod_{b} \stackrel{I_b}{\pi}{}\!\!^{j_b}_{k_b} (\Lambda_0^{-1} U_b \Lambda_0)\;
\prod_{e} \stackrel{I_e}{\pi}{}\!\!^{j_e}_{k_e} (\Lambda_0^{-1} U_e \Lambda_e)\;.
\end{eqnarray}
Finally, the invariance under the action of ${\cal C}$ on each (spinless) particle but the observer implies directly that
$k_e=0$ for all $e=1,\cdots, n-1$. Yet $\stackrel{I}{\pi}{}\!\!^j_0$ are functions on the sphere $G/U(1)$ (spherical
harmonic functions) and we can write them as follows:
\begin{eqnarray}
\forall \; g \in SU(2) \;\;\; \stackrel{I}{\pi}\!\!{}(g)^j_0 \; = \; \int dh \; \stackrel{I}{\pi}{}\!\!(gh)^j_0 \;\;\;
\text{where $h \in U(1)$}\;. 
\end{eqnarray}
The space of such invariant functions will be denoted $\text{Cyl}_{Pinv}(\gamma;G)$. It
inherits the measure from $\text{Cyl}(\gamma;G)$ due to left and right invariance
of the Haar measure and then it possesses a natural pre-Hilbert structure. Its completion is the partial kinematical
Hilbert space ${\cal H}_{Pkin}(\gamma;G)$ and it is a sub-Hilbert space of ${\cal H}(\gamma;G)$. 

Moreover, it is a straightforward to show that any element of $Fun(G^{2g} \otimes (S^2)^{n-1})$ is a kinematical partial Hilbert
space. 
Finally, the proposition (\ref{Pkin}) follows immediately. $\Box$

\medskip

Let us propose several remarks concerning the partial kinematical Hilbert space. 

1. First of all, note that one can precise the proposition in the sense that the isomorphism holds when the partial kinematical 
Hilbert space is defined with elements $\psi$ which are distributional for the observer variable and functional for the others.
To clarify this claiming, let us illustrate it in the case of the sphere with particles.
Indeed, the fact that the observer is colored with an intertwiner means that any spin-network states can be written as:
\begin{eqnarray}
\psi(A) \; = \; \int dg \; f(\Lambda_0 g, \Lambda_1 g,\cdots,\Lambda_{n-1}g) \;.
\end{eqnarray}
It is then clear that to make $\Psi(A)$ a function on $(S^{2})^{n-1}$, it is enought to require that $f$ is a distribution on
the first argument and a function for the others. We made implicitely this asumption in the proof of the proposition.

2. As a secong remark, we underline the fact that the proposition (\ref{Pkin}) is valid whatever the choice of the observer among the
different particles. Indeed, each partial kinematical Hilbert space are isomorphic one to the other. 
In fact, the previous proposition is still valid when the base point $x$ is not a point of $\partial \Sigma$. 
To understand this point, we recall that ${\cal H}_{Pkin}(\gamma;G)$ is a sub-Hilbert space of ${\cal H}(\gamma;G)$ and 
there exists a projector 
$P_{Pkin}:{\cal H}(\gamma;G) \rightarrow {\cal H}_{Pkin}(\gamma;G)$ given by a product $P_{Pkin}=P_G \cdot P_{\cal C}$ defined by:
\begin{eqnarray}
(P_G \psi)(A) & = & \int dy \prod_{j=0}^{n-1} dy_j\; 
f(\bigotimes_{a,b=1}^g y^{-1}U_a y \otimes y^{-1}U_b y \otimes \bigotimes_{i,e=0}^{n-1} y^{-1}U_ey_e \otimes y_i^{-1}\Lambda_i) \\
(P_{\cal C} \psi)(A) & = & \int \prod_{j=0}^{n-1} dh_j \; f(\bigotimes_{a,b=1}^g U_a\otimes U_b \otimes 
\bigotimes_{i,e=0}^{n-1} U_e \otimes \Lambda_i h_i) \;,
\end{eqnarray}
for any $\psi \in {\cal H}(\gamma;G)$ associated to a function $f \in Fun(G^{2(g+n)})$. 
Note that $dy$ or $dy_i$ is the Haar measure on the group $G$ and $dh_i$ the Haar measure on the group $S^1$. Thus, 
any state $\widetilde{\psi} \in {\cal H}(\gamma;G)$ is a partial kinematical state if there exists a state $\psi \in {\cal H}(\gamma;G)$
associated to a function $f \in Fun(G^{2(g+n)})$ such that:
\begin{eqnarray}
\widetilde{\psi}(A) \; = \; (P \psi)(A) \; = \; (P_G\cdot P_{\cal C} \; \psi)(A)\;.
\end{eqnarray}
Therefore, after some calculations it is direct to see that there exists $j \in [0,n-1]$ such that:
\begin{eqnarray}
\widetilde{\psi}(A) \; = \; \tilde{f}(\bigotimes_{a,b=1}^g \Lambda_j^{-1} U_j^{-1} U_a U_j \Lambda_j \otimes 
\Lambda_j^{-1} U_j^{-1} U_b U_j \Lambda_j \otimes \bigotimes_{e \neq j} \Lambda_j^{-1} U_j^{-1} U_e \widetilde{\Lambda}_e)
\end{eqnarray}
where the function $\tilde{f} \in Fun(G^{2g} \otimes (S^2)^{n-1})$ is related to the function $f$ by the expression:
$$
\tilde{f}(\bigotimes_{a,b=1}^g U_a \otimes U_b \otimes \bigotimes_{e\neq j} U_e) \; \equiv \; 
\int \! dy \prod_{j=0}^{n-1} dy_j \; dh_j $$
$$
f(\bigotimes_{a,b=1}^g y^{-1} (U_j \Lambda_j)^{-1} U_a U_j \Lambda_j y \otimes y^{-1} (U_j \Lambda_j)^{-1}U_b U_j \Lambda_j y \otimes 
\bigotimes_{e,i=0}^n y^{-1}(U_j \Lambda_j)^{-1}U_e\Lambda_e h_e \otimes 1 ).$$
The choice of $j$ corresponds to the choice of the observer. Therefore, the proposition (\ref{Pkin}) is valid for any choice of the 
base point. In the sequel, we will assume that the observer is the particle ${\cal P}_0$.

3. The final remark is that one can recover the kinematical
Hilbert space from the partial kinematical Hilbert space as follows:
\begin{eqnarray}
{\cal H}_{kin}(\gamma;G) \; = \; {\cal H}_{Pkin}(\gamma;G)/S^1
\end{eqnarray}
where $S^1$ is the symmetry group which acts by right multiplication on the variable $\Lambda_0$ (\ref{calC}).

\medskip

The next step is the construction of the partial physical Hilbert space ${\cal H}_{Pphys}(\gamma;G)$. For that purpose, we start by
defining the extractor $P: \text{Cyl}(\gamma;G) \rightarrow \text{Cyl}^*(\gamma;G)$ formally defined by the expression:
\begin{eqnarray}
\forall \;\; \varphi, \; \psi \; \in \; \text{Cyl}(\gamma;G) \;,\;\;\;\;\;
P(\psi)(\varphi) \; \equiv \;  \int d\mu[A] \; \overline{\psi(A)} \; {\cal K}(A) \; \varphi(A) \;.
\end{eqnarray}
The measure on the space of discrete connections ${\cal A}_\gamma$ is constructed from the $SU(2)$ Haar measure $d\mu$ via the 
relation $d\mu[A] = d\mu^{\otimes 2(g+n)}$ for ${\cal A}_\gamma$ is isomorphic to $2(g+n)$ copies of $SU(2)$. The Kernel ${\cal K}$
is a distribution defined by:
\begin{eqnarray}
{\cal K}[A] \; \equiv \; \delta(\Lambda_0 h(m_0) \Lambda_0^{-1} \prod_{e=1}^{n-1} U_e \Lambda_e h(m_e) \Lambda_e^{-1} U_e^{-1} 
\prod_{(a,b)}[U_{a},U_{b}]) \;.
\end{eqnarray}
We have introduced the usual notation $[U_a,U_b]=U_aU_bU_a^{-1}U_b^{-1}$; $(a,b)$ denotes the set of the pairs of non-contractible loops;
and the elements $h(m_e)$ are in the Cartan (diagonal) torus of $SU(2)$ 
fixed by the mass $m_e$ by $h(m_e)=\text{diag}(e^{+i m_e},e^{-i m_e})$ in the $SU(2)$ fundamental representation. 

The extractor $P$ defines a bilinear form on the space $P(\text{Cyl}(\gamma;G)) \subset \text{Cyl}(\gamma;G)^*$; it is denoted
$<;>$ and is given by:
\begin{eqnarray}\label{Hform}
\forall \; \psi,\; \varphi \; \in \; \text{Cyl}(\gamma;G) \;,\;\;\;\; <P(\psi);P(\varphi)> \; \equiv \; P(\psi)(\varphi) \;. 
\end{eqnarray}
In fact, this bilinear form defines a scalar product for it satisfies the following properties:
\begin{enumerate}
\item It is definite positive: the kernel $\cal K$ is a delta distribution on the group and therefore can be obtained as a limit
of positive functions on $G$.
\item It is convergent as soon as the stability condition $2g+n-1>0$ is satisfied; we have the following bound:
\begin{eqnarray}\label{bound}
<P(\varphi),P(\psi)> \; \leq \; \vert\!\vert \varphi \vert\!\vert \cdot \vert\!\vert \psi \vert\!\vert 
\frac{1}{\prod_{e=0}^{n-1}\sin (m_e)}
\sum_{k=1}^{\infty} \frac{\prod_{e=0}^{n-1}\sin(k m_e)}{k^{2g+n-2}} \;.
\end{eqnarray}
We have introduced the notation $\vert\!\vert\cdot\vert\!\vert$ for the $L^2$ norm on the group $SU(2)$.
The bound (\ref{bound}) is a direct consequence of the Plancherel theorem for $SU(2)$ and of the expression of the $SU(2)$ character 
in any $I \in \frac{1}{2} \mathbb N$ irrep, i.e. $\chi_I (h(m_e))=\sin((2I+1)m_e)/\sin(m_e)$.
\end{enumerate}

The vector space of partial physical states is defined by $P(\text{Cyl}_{Pinv}(\gamma;G))$. It is clear that the kernel ${\cal K}$
is in fact a distribution on the space $\text{Cyl}_{Pinv}(\gamma;G)$ and therefore $P(\text{Cyl}_{Pinv}(\gamma;G)) \subset 
\text{Cyl}_{Pinv}(\gamma;G)^*$. In particular, the kernel $\cal{K}$ is co-invariant under the transformations 
$\Lambda_e \mapsto \Lambda_e h_e$ for any $e$ and $h_e \in U(1) \subset SU(2)$. Thus, $\cal{K}$ depends only on the equivalent
classes $\tilde{\Lambda}_e \equiv \int dh_e \; \Lambda_e h_e \in S^2$ where $dh_e$ is the $U(1)$ Haar measure. 
For convenience, it will be useful to introduce the notation $d\tilde{\Lambda}$ for the normalized measure on the 2-sphere.

The bilinear form (\ref{Hform}) defines a pre-Hilbert structure on $P(\text{Cyl}_{Pinv}(\gamma;G))$ whose completion 
(up to null-vectors) is the
partial physical Hilbert space ${\cal H}_{Pphys}(\gamma;G)$. When restricted on ${\cal H}_{Pphys}(\gamma;G)$, the scalar product
(\ref{Hform}) is called the partial physical scalar product and is denoted $<;>_{Pphys}$.

\begin{proposition}[Partial physical Hilbert space]\label{PHilbert}
Let $f \in Fun(G^{2g} \otimes (S^2)^{n-1})$. One defines a partial physical states $P(\psi)$
where $\psi \in \text{Cyl}_{Pinv}(\gamma;G)$ was introduced in the proposition (\ref{Pkin}).
Therefore, there is a map 
$$ 
Fun(G^{2g} \otimes (S^2)^{n-1}) \; \longrightarrow \; {\cal H}_{Pphys}(\gamma;G)\;.
$$
When $Fun(G^{2g} \otimes (S^2)^{n-1})$ is endowed with the following Hilbert structure:
$$
\forall \; f,g \; \in Fun(G^{2g} \otimes (S^2)^{n-1}) 
$$ 
$$
(f,g) \; \equiv \; \int dadb \;
d\tilde{z} \; \overline{f(a,b,\tilde{z})} \; \delta(h(m_0) \prod_{e=1}^{n-1} {z}_e h(m_e) {z}_e^{-1} 
\prod_{i=1}^g [a_i,b_i]) \; g(a,b,\tilde{z})\;,
$$
the previous map is an isometry.
Note that we have introduced the notation $a=(a_i)_i$, $(b_i)_i$, $z=(z_e)_e$ and $da=\prod_{i} da_i$, $db=\prod_{i} db_i$
for the measures.
\end{proposition}

{\it Proof.}

The proof of the proposition (\ref{Pkin}) tells us that any invariant cylindrical function $\psi$ is completely caracterized by a function 
$f \in Fun(G^{2g} \otimes (S^2)^{n-1})$ (to be more precise, the vector space $\text{Cyl}_{Pinv}(\gamma;G)$ corresponds to the set
of polynomial functions on $G^{2g} \otimes (S^2)^{n-1}$ and after completion with respect to the kinematical measure one sees 
$\text{Cyl}_{Pinv}(\gamma;G)$ as the set of functions on $G^{2g} \otimes (S^2)^{n-1}$). 

Yet, by definition, the partial physical Hilbert space ${\cal H}_{Pphys}(\gamma;G)$
is the image of $Fun(G^{2g} \otimes (S^2)^{n-1})$ by the operator $P$ and then any element of ${\cal H}_{Pphys}(\gamma;G)$ 
can be written as ${\cal K}[A] \psi(A) \in \text{Cyl}_{Pinv}(\gamma;G)^*$. It is then natural to identify ${\cal H}_{Pphys}(\gamma;G)$
with the space $Fun(G^{2g} \otimes (S^2)^{n-1})$.

The last point of the proposition is a direct consequence of the right/left invariances of the $SU(2)$ Haar measure. 
$\Box$

\medskip

Let us finish this section with some remarks.

Remark 1. The isometry given in the previous proposition holds for any choice of minimal graph $\gamma$ and any choice of observer. 
Each partial physical Hilbert space ${\cal H}_{Pphys}(\gamma;G)$ where $\gamma$ is any minimal graph 
are isomorphic one to the other and therefore are equivalent. For that reason,
we will use the notation ${\cal H}_{Pphys}^{g,n}([m];G)$
to denote the partial physical Hilbert space where $[m]=(m_0,\cdots,m_{n-1})$.
We have emphasized the explicit characteristics of the partial physical Hilbert space:
the topological structure 
of $\Sigma$,  the colors of the boundaries, i.e. the
masses of the particles and the gauge group $G$ of course. 

\medskip

Remark 2. The order of the particles matters in the definition of ${\cal H}_{Pphys}^{g,n}([m];G)$.
The space ${\cal H}_{Pphys}^{g,n}([m];G)$ is completely and uniquely caracterised by a given flat graph
of the type (figure \ref{minimalgraph}) where there is an order between the links. If one modifies the order between the links,
then the definition the structure of the partial physical Hilbert space is slightly modified.
However, two different orders define isomorphic partial physical Hilbert space. For instance, if we consider a sphere $\Sigma=S^2$ 
with three particles associated to the masses $(m_0,m_1,m_2)$, then the following map is an isometry:
\begin{eqnarray}
{\cal H}_{Pphys}^{0,3}(m_0,m_1,m_2;G) & \longrightarrow & {\cal H}_{Pphys}^{0,3}(m_0,m_2,m_1;G) \nonumber \\
f & \longmapsto & g: (\tilde{z}_1,\tilde{z}_2) \mapsto g(\tilde{z_1},\tilde{z_2}) = f(\tilde{z_1},z_1h(m_1)^{-1}z_1^{-1}\tilde{z}_2)\;. 
\end{eqnarray}
It is indeed easy to verify that $\vert \! \vert f \vert \! \vert _{Pphys} = \vert \! \vert g \vert \! \vert _{Pphys}$.
One can establish a more general property that we will discuss in the next section.

\medskip

Remark 3. The physical Hilbert space ${\cal H}_{phys}^{g,n}([m];G)$ is easily obtained as the coset:
$${\cal H}_{phys}^{g,n}([m];G) \; \equiv \; {\cal H}_{Pphys}^{g,n}([m];G)/S^1\;.$$ 
The action of $S^1$ on ${\cal H}_{Pphys}^{g,n}([m];G)$ is naturally defined by:
\begin{eqnarray}\label{S1action}
S^1 \times {\cal H}_{Pphys}^{g,n}([m];G) & \longrightarrow & {\cal H}_{Pphys}^{g,n}([m];G) \\
(h,f) &  \longmapsto & f^{h}\; : \; y,\tilde{z} \; \mapsto \; f^h(a,b,\tilde{z}) 
\; = \; f(h^{-1}ah,h^{-1}bh,h^{-1}\tilde{z})\;. 
\end{eqnarray}
We have used the notations of the proposition (\ref{PHilbert}) with $h^{-1}ah=(h^{-1}a_i h)_i$ and so on.
Therefore, the physical Hilbert space is obviously a sub-Hilbert space of ${\cal H}_{Pphys}^{g,n}([m];G)$ and can be obtained as the 
image of the following projector:
\begin{eqnarray}
{\cal H}_{Pphys}^{g,n}([m];G) \; \longrightarrow \; {\cal H}_{phys}^{g,n}([m];G) \;\;\;\;\;\;\; f \; \longmapsto \tilde{f}(a,b,\tilde{z})
\; = \; \int dh \; f(h^{-1}ah,h^{-1}bh,h^{-1}\tilde{z})\;.
\end{eqnarray}
This is in fact a trivial application of refined algebra techniques in the compact $S^1$ case. The Hilbert structure of 
${\cal H}_{phys}^{g,n}([m];G)$ is defined from this map as the kernel ${\cal K}$ is invariant by the action (\ref{S1action}).

\medskip

Remark 4. From the very definition of the partial physical Hilbert space as a subset of $\text{Cyl}(\gamma;G)^*$, 
${\cal H}_{Pphys}(\gamma;G)$ is in fact a subset of  $\mathbb C[G]^{2g} \otimes  \mathbb C[S^2]^{n-1}$ where 
$\mathbb C[G]$ is the group algebra and $\mathbb C[S^2]=\mathbb C[G]/U(1)$ is called the 2-sphere algebra. The group algebra
is defined as the algebra of formal sum of group elements. As $G$ is compact there is a trivial isomorphism which identifies
the spaces $\mathbb C[G]^{2g} \otimes  \mathbb C[S^2]^{n-1}$ and $Fun(G^{2g}\otimes (S^2)^{n-1})$.

\medskip

Remark 5. Let us recall that any particle $e \in [0,n-1]$ is classically completely caracterized by its mass
$m_e$ and the direction of its momentum given by $\Lambda_e$. The particle ${\cal P}_0$ plays the role of the observer and the
variable $\Lambda_0$ does not appear anymore in the definition of partial physical states. One can interpret this observation by the
fact that we work in the rest-frame of the observer and then $\Lambda_0$ is set to the identity: the momenta of the other particles
are defined in that frame.

Besides, one can caracterized the particle $e$ by an element $k_e \in G$ such that $k_e = \Lambda_e h(m_e)\Lambda_e^{-1}$, i.e. $k_e$
is an element of the conjugacy class of the element $h(m_e)$, denoted $C(m_e)$:
\begin{eqnarray}
C(m_e) \; \equiv \; \{k\in SU(2) \; \vert \; \exists \; y \in \; SU(2) \;, \; \text{such that} \;\; k=yh(m_e)y^{-1}\}\;.
\end{eqnarray}
Thus, any element of ${\cal H}_{Pphys}^{g,n}([m];G)$ 
can be viewed as an element of $Fun(G^{2g}\otimes \bigotimes_{e=1}^{n-1}C(m_e))$. Using the Kirillov formulae for the $SU(2)$ Haar 
measure,
\begin{eqnarray}
dk_e \; = \; \frac{1}{\pi} \sin^2(m_e) \; dm_e \; d\tilde{\Lambda}_e,
\end{eqnarray}
the partial physical scalar product between two states $\psi$ and $\varphi$ respectively associated to the functions $f$ and $g$ 
is then given by:
\begin{eqnarray}
<\psi;\varphi>_{Pphys} & = & (f,g)  \\
& = & \frac{\pi^{n-1}}{\prod_{e=1}^{n-1} \sin^2(m_e)} \; 
\int dy \; dk \; \overline{f(a,b,k)} \; \delta(h(m_0)\prod_{e=1}^{n-1} k_e \prod_{i=1}^g[a_i,b_i])\; g(a,b,k).\nonumber
\end{eqnarray}
We recall that the functions $f$ and $g$ are non-null if $g_e \in C(m_e)$. This reformulation of the partial physical Hilbert space
will be convenient in the next section.

\medskip

Remark 6. The last remark concerns the normalization of the partial physical scalar product.
whose is defined only up to a constant $\lambda(g,n,[m])>0$.
The constant depends a priori on the genus $g$ of 
$\Sigma$, the number of particles $n$ and the masses of the particles.  
By a direct calculation, one shows that the squared norm of the identity function $e^{g,n}[m] \in 
{\cal H}_{Pphys}^{g,n}([m];G)$ is given by the volume of the space of 
flat $SU(2)$ connections on the punctured surface $\Sigma$, i.e.:
\begin{eqnarray}\label{Unitnorm}
\vert \! \vert e^{g,n}[m] \vert \! \vert_{Pphys}^2 \; = \; \lambda(g,n,[m])\;
\sum_{k=1}^{+\infty}\; {k^{2-2g-n}}\; \prod_{e=0}^{n-1} \frac{\sin(km_e)}
{\sin(m_e)} \; .
\end{eqnarray} 
At this point, there is not physical reason to impose
the value of $\lambda$. But, we can show that the value of $\lambda(g,n,[m])$ is completely determined by the 
coefficients $\lambda(0,3,[m])$ and the problem to fix it reduces to the fixation of the normalization factor in the (minimal) case
of the sphere with 3 punctures. The method is similar to the one proposed in \cite{Alekseev} and is given in the appendix B:
if we assume that $\lambda(0,3,[m])=\lambda$ is a constant independent of the masses, then we show that 
$\lambda(g,n,[m])=\lambda^{n+2g-2}$.

\section*{3. Transition amplitudes and the Quantum double}
In this section, we are going to show how the Drinfeld double $DSU(2)$ structure appears in the context of three dimensional Riemannian 
gravity coupled to massive point particles: the partial physical Hilbert space is in fact isomorphic as a Hilbert space to a tensor
product of simple representations of $DSU(2)$. In that picture, the partial physical scalar product between two states is an 
intertwining coefficient between simple representations. Thus, the so-called Barrett-Crane (BC) intertwiner \cite{BC1} 
plays a crucial role in three dimensional quantum gravity whereas it was introduced as a model of four dimensional gravity.

This section is organized as follows. First, we recall basic facts concerning the Drinfeld double $DSU(2)$: definition,
representation theory. We also introduce the notion of simple representation and symmetric (or BC) intertwiner. Then,
we show the role of $DSU(2)$ in the case of a sphere with $n$ particles: $DSU(2)$ appears as the symmetry group of the
quantum theory and the non-trivial braiding is interpreted in the language of LQG. Finally, we generalize the previous results
in the case of any Riemann surface $\Sigma$.

\subsubsection*{3.1. The quantum double $DSU(2)$: definition and properties}
The general definition of the quantum double (or the Drinfeld double) of a Hopf algebra $A$ is briefly recalled in 
the appendix A. The general definition is then illustrated in the generic case of a finite group \cite{DPR}. In this section,
we consider the case of the compact group $SU(2)$, i.e. the case where $A=\mathbb C[SU(2)]$ \cite{KM}.

The Drinfeld double $DSU(2)=\mathbb C[SU(2)] \otimes F(SU(2))^{op}$ is a Hopf algebra whose definition is precisely given in
the appendix. In particular, it is quasi-triangular and admits the group algebra $\mathbb C[SU(2)]$ and the algebra of functions
$F(SU(2))^{op}$ (with opposite co-product) as sub-Hopf algebra.

\medskip

{\it 3.1.1. Hopf algebra structure}

\medskip

The Hopf algebra structure of $\mathbb C[SU(2)]$ is defined by the group law and the following co-algebra relations:
\begin{eqnarray}
\forall x \in SU(2),\;\; \Delta(x)=x\otimes x,\;\; S(x)=x^{-1},\;\; \epsilon(x)=\delta(x)\;.
\end{eqnarray}
Note that there exists a right and left invariant Haar measure on $\mathbb C[SU(2)]$ given by:
\begin{eqnarray}
h:\mathbb C[SU(2)] \; \longrightarrow \; \mathbb C \;\;\; \text{and} \;\;\; h(x)=\delta(x)\;.
\end{eqnarray}
The algebra of functions $F(SU(2))$ is commutative and its coalgebra struture is given by:
\begin{eqnarray}
\Delta(f)(x,y)\;=\; f(xy),\;\; S(f)(x)=f(x^{-1}),\;\; \epsilon(f)=f(1)\;,
\end{eqnarray}
where $f$ is a function and $x,y \in G$. We know that $F(SU(2))$ admits a Haar measure that we have denoted
$\int \! dx$ as usual.

Finally, the Drinfeld double is defined by the previous formulae and the following ones (which are a direct application of the 
defintion given in the appendix):
\begin{eqnarray}
(x\otimes f)(y \otimes g) \; = \; xy \otimes (f \circ Ad_y) g \\
\Delta(x\otimes f)(a,b)=f(ba) x\otimes x \;,
\end{eqnarray}
where $f,g \in F(SU(2))$, $x,y,a,b \in SU(2)$. The action of the antipode reads 
\begin{eqnarray}
S(x\otimes f)(a)=x^{-1}\otimes f(x^{-1}a^{-1}x)
\end{eqnarray}
for any $a \in G$ and the co-unit is simply given by $\epsilon(x\otimes f)=f(1)$.

\medskip

{\it 3.1.2. Representation theory}

\medskip

Unitary irreducible representations of $DSU(2)$ have been studied in \cite{KM} and are classified by a couple $(m,s)$:
$m\in[0,2\pi]$ labels a conjugacy class of $SU(2)$, denoted $C(m)$, whose representative is 
still chosen to be $h(m)$; 
$s\in \mathbb Z$ is an integer which labels irreducible representations of the centralizer $Z(m)$
of the element $h(m)$.
Note that $Z(m) \simeq U(1)$ 
for $m\neq 0, 2\pi$; otherwise $Z(m)$ is obviously the group $G=SU(2)$ itself. In the generic case $m\neq 0,2\pi$, the vector 
space of a representation $(m,s)$ consists in the subspace of functions on $SU(2)$ defined by:
\begin{eqnarray}
{\cal V}_{m,s} \; \equiv \; \{\varphi:SU(2) \rightarrow \mathbb C \vert \forall \; a \in SU(2),\; \varphi(ah_m)=e^{ism}\varphi(a)\}\;. 
\end{eqnarray}
The representation of any element $(x\otimes f) \in DSU(2)$ on the above vector space is defined by:
\begin{eqnarray}
(\pi_{m,s}(x\otimes f)\varphi)(a) \; = \; f(ah_ma^{-1})\; \varphi(x^{-1}a)\;.
\end{eqnarray}
The vector space ${\cal V}_{m,s}$ inherits a natural Hilbert space structure. 
The particular case $m \in \{0,2\pi\}$ has been considered in \cite{KM}. In the sequel, we use the notation 
$\overline{m}=2\pi - m$.

\medskip

{\it 3.1.3. Simple representations and symmetric intertwiners}

\medskip

Representations of the form $(m,0)$ are called simple representations \cite{FNR} and are the building blocks of the Barrett-Crane model
\cite{BC1}.
The vector space of a simple representation is isomorphic to the space of functions on the sphere $S^2=SU(2)/U(1)$.
Any simple representation admits a normalized $SU(2)$-invariant vector denoted $\omega=1$. 
The vector space associated to a simple representation $m$ will be denoted ${\cal V}_m$ (we omit the subscript $0$ for the spin).
It is obvious that ${\cal V}_m \simeq Fun(S^2)$ and we will identify in the following the spaces ${\cal V}_m$ and the conjugacy
class $C(m)$ which contains the information on the mass. We will use the same notation $\psi(\tilde{x})$ or $\psi(xh(m)x^{-1})$ 
for the state $\psi$ when viewed as an element of $Fun(S^2)$ or an element of $C(m)$.
 
The notion of symmetric intertwiner (or the Barrett-Crane intertwiner) has been introduced a long time ago \cite{BC1,FNR} and is 
defined as an intertwiner between simple representations whose decomposition into three-valent intertwiners introduces only simple
representations in the intermediate channel. Up to a normalization, the simple intertwiner is unique \cite{Reis}.

There exists an integral formulation of the symmetric intertwiner.
To present this formulation, we need to recall the definition of the symmetric propagator \cite{FNR}.

Given a simple representation
labelled by $m$, the symmetric kernel $K_m$ is an element of $F(SU(2))^{\star} \subset DSU(2)^\star$ defined by
\begin{eqnarray}
K_m\;: \; F(SU(2)) \longrightarrow \mathbb C\;,\;\; f \longmapsto \; K_m(f) \; = \; \int dx \; f(xh(m)x^{-1})\;.
\end{eqnarray}
It is convenient and equivalent to view (by duality) the symmetric kernel as an element of $\mathbb C[SU(2)]$ as follows:
\begin{eqnarray}
K_m \; = \; \int dx \; {\mathbb I}_m(x) \; x \;\;\;\;\;\;\; \text{with} \;\;\; {\mathbb I}_m(x) = \int dy \; \delta(x y h(\overline{m})
y^{-1}) \;.
\end{eqnarray}
Note that $\mathbb I_m$ is the characteristic function of the conjugacy class $C(m)$ and can be decomposed into Fourier modes as follows:
\begin{eqnarray}
\mathbb I_m(x) \; = \; \sum_{I \in \frac{1}{2} \mathbb N} \chi_I(m)\;\chi_I(x) \;\;\;\;\;\;\;\; \text{with} \;\;\;\; 
\chi_I(m) \equiv \chi_I(h(m))\;.
\end{eqnarray}
If $K_m$ had been a function on a ``classical'' group $G$, we would have defined the ``classical'' propagator as the function 
$K(x,y)\equiv K(xy^{-1}) \in F(G)^{\otimes 2}$. In our case, the propagator, denoted $K_m(x,y)$, is an element of 
$\mathbb C[SU(2)]^{\otimes 2}$ defined by:
\begin{eqnarray}
K_m(x,y) \; \equiv \; (1 \otimes S)\Delta(K_m) \; = \; \int dx \; \mathbb I_m(x) \; x\otimes x^{-1} \;.
\end{eqnarray}
This propagator is the building block of the symmetric intertwiner whose construction is given next \cite{FNR}.
Let $[p]=(p_1,\cdots,p_a)$ and $[q]=(q_1,\cdots,q_b)$ two ordered families of simple representations. 
One defines the symmetric intertwiner between the representations $[p]$ and $[q]$,
\begin{eqnarray*}
\iota_s([p];[q]) \; : \; \bigotimes_{i=1}^a {\cal V}_{p_i} \; \longrightarrow \; 
\bigotimes_{j=1}^b {\cal V}_{q_j} \;,
\end{eqnarray*}
by its matrix elements given by:
\begin{eqnarray}\label{matelements}
<\bigotimes_{\ell=1}^b \varphi_l \vert \iota_s([p];[q]) \bigotimes_{k=1}^a \psi_k> \; = \;  h \left(
\prod_{j=1}^b  (\overline{\varphi_j} \otimes 1) K_{q_j}(x_j,g) \;  
\prod_{i=a}^1  (\psi_i \otimes 1)K_{\overline{p_i}}(y_i,g)\right)\;,
\end{eqnarray}
where $\psi_i \in {\cal V}_{p_i}$ and $\varphi_j \in {\cal V}_{q_j}$ for any $i$ or $j$. Note that $h$ is the Haar measure on the
group algebra $\mathbb C[SU(2)]$ and the products in the formulae (\ref{matelements}) are ordered such that when one develops the 
expression, one obtains:
\begin{eqnarray}
<\bigotimes_{\ell=1}^b \varphi_\ell\vert \iota_s([p];[q]) \bigotimes_{k=1}^a \psi_k>  =  \int \! \prod_{i,j} dx_j \; dy_i \;
\mathbb I_{q_j}(x_j)\overline{\varphi_j(x_j)}\; \delta(\prod_{\ell=1}^b x_\ell^{-1} \prod_{k=a}^1 y_k)\;\mathbb I_{p_i}(y_i)\psi_i(y_i) \;.
\end{eqnarray}

\medskip

Let us propose several remarks concerning the symmetric intertwiner.

Remark 1.
If one views the states $\psi_i$ and $\varphi_j$ as functions on the 2-sphere, the previous formulae reduces to the following one:
\begin{eqnarray}
<\bigotimes_{\ell=1}^b \varphi_\ell\vert \iota_s([p];[q]) \bigotimes_{k=1}^a \psi_k>  \; = \;  \int \; \prod_{j=1}^b 
 \overline{\varphi_j(\tilde{x}_j)} \; d\tilde{x}_j \; \prod_{i=1}^a  
\psi_i(\tilde{y}_i) \; d\tilde{y}_i 
 \; \delta(\prod_{l=1}^b x_l h(\overline{q_l})x_l^{-1} \prod_{k=a}^1 y_k h(p_k) y_k^{-1})\;\nonumber
\end{eqnarray}
In the particular (and singular) case where $\varphi_\ell$ and $\psi_k$ are picked at the values $\tilde{x}_\ell$
and $\tilde{y}_k$ on the sphere, the symmetric intertwiner is a distribution given by:
\begin{eqnarray}
<\bigotimes_{\ell=1}^b \tilde{x}_\ell \vert \iota_s([p];[q]) \bigotimes_{k=1}^a \tilde{y}_k>  \; = \;
 \delta(\prod_{\ell=1}^b x_\ell h(\overline{q_\ell}) x_\ell^{-1}
\prod_{k=a}^1 y_k h(p_k) y_k^{-1})\;.
\end{eqnarray}

\medskip

Remark 2.
We easily show that the decomposition of the symmetric intertwiner introduces only simple representations in
the intermediate channel. Using the previous notations, we have for instance:
\begin{eqnarray}
\iota_s([p];[q]) = \frac{1}{\pi}\int dr \; P(r) \; \iota_s(p_1,p_2;r) \; \circ \;\iota_s(\overline{r},p_3,\cdots,p_a;[q])\;,
\end{eqnarray}
where $\circ$ denotes the usual composition between intertwiners and $P(r)=\frac{\sin^2 r}{\pi}$ is the Plancherel measure.

\medskip

Remark 3.
Therefore, we concentrate for the moment on the three valent intertwiner $\iota_s(p;q_1,q_2): {\cal V}_p 
\rightarrow {\cal V}_{q_1}\otimes {\cal V}_{q_2}$. We start by saying that it is indeed an intertwiner, i.e. it satisfies the
following identity:
\begin{eqnarray}
\iota_s(p;q_1,q_2) (\pi_{p} (z\otimes f)) \; = \; (\pi_{q_1} \otimes \pi_{q_2})\Delta(z\otimes f) \; \iota_s(p;q_1,q_2)\;
\end{eqnarray}
for any $z\otimes f \in DSU(2)$. 
The symmetric intertwiner is not normalized and we define the coefficient $\Theta(p,q_1,q_2)$
such that the intertwiner $\Theta(p,q_1,q_2)^{-1/2} \iota_s(p;q_1,q_2)$ is normalized, i.e:
\begin{eqnarray}
\frac{1}{\sqrt{\Theta(p,q_1,q_2) \Theta (q_1,q_2,p')}} \iota_s(p;q_1,q_2) \circ \iota_s(q_1,q_2;p') \; = \; \frac{\delta(p-p')}{P(p)}
id_{{\cal V}_p}\;.
\end{eqnarray}
A straightforward calculation shows that:
\begin{eqnarray}
\Theta(p,q_1,q_2) \; = \; 
<1 \otimes 1 \vert \iota_s([p];[q]) \vert 1> \; = \; \frac{\pi}{4} \frac{Y(p,q_1,q_2)} {\sin p \; \sin q_1 \; \sin q_2} \;
\end{eqnarray}
where $Y$ is the characteristic function of the set $\{(a,b,c)\vert a\leq b+c, \; b \leq c+a,\; c \leq a+b\}$.

\medskip

Remark 4. From the previous remarks, it is easy to show that $\iota_s([p];[q])$ is an intertwiner, i.e. it satisfies:
\begin{eqnarray}
\iota_s([p];[q]) \; (\otimes_{i=1}^a \pi_{p_i})\Delta^{(a)}(z \otimes f) \; = \; (\otimes_{i=1}^b \pi_{q_i})\Delta^{(b)}(z \otimes f)\;
\iota_s([p];[q])\;.
\end{eqnarray}
We used the notation $\Delta^{(n)}$ defined by $\Delta^{(1)} = \Delta$ and $\Delta^{(n)}= (id \otimes \cdots \otimes id 
\otimes \Delta) \Delta^{(n-1)}$. 

\medskip

Remark 5. It is straightforward to extend the action of the symmetric intertwiner $\iota_s([p];[q])$ to the space 
$Fun((S^2)^{\times a})$.

\medskip

{\it 3.1.4. R-matrix and braiding}

\medskip

Another important property of the quantum double is its quasitriangularity.
Indeed, the quantum double is, by construction, a quasitriangular Hopf-algebra and therefore admits a $R$-matrix. 
The universal expression of $R$ is given by the formula:
\begin{eqnarray}
R \; = \; \int dg \; (g \otimes 1) \otimes (1 \otimes \delta_g)\;.
\end{eqnarray}
Then, we can evaluate the $R$-matrix on any pair of representations. We are particularly interested in the case of simple representations.
Let $(p,q)$ be a pair of simple representations, the evaluation of $R$ in these representations is given by:
\begin{eqnarray}
R_{p,q} \; \equiv \; \pi_{p} \otimes \pi_{q}(R) \; : \; {\cal V}_{p} \otimes {\cal V}_{q} \; \longrightarrow
{\cal V}_{q} \otimes {\cal V}_{p}
\end{eqnarray}
such that, for any $\psi,\varphi \in Fun(S^2)$, then:
\begin{eqnarray}\label{Rmatrix}
R_{p,q} (\psi \otimes \varphi) (\tilde{x},\tilde{y}) \; = \; \varphi(x h(\overline{p}) x^{-1} \tilde{y}) \; \psi(\tilde{x})
\end{eqnarray}
where $g\tilde{x}$ means the action of $g \in SU(2)$ on the point $\tilde{x}\in S^2$. 
The evaluation of the inverse $R$-matrix on simple representations is given by: 
\begin{eqnarray}\label{Rinverse}
R^{-1}_{p,q} (\psi \otimes \varphi) (\tilde{x},\tilde{y}) \; = \; \varphi(\tilde{y}) \; \psi(yh(q)y^{-1} \tilde{x})\;.
\end{eqnarray}

It is convenient to write the action of the R-matrix and of its inverse on the states viewed as functions on the conjugacy
class. In that context, we do not need to specify the evaluation representation of the $R$ matrix (and its inverse $R^{{}_{-1}}$) 
and we have:
\begin{eqnarray}
R(\psi \otimes \varphi) (x,y) \; = \; \varphi(x^{-1}yx) \psi(x) \;\;\; \text{and} \;\;\; 
R^{-1}(\psi \otimes \varphi) (x,y) \; = \; \varphi(y) \psi(yxy^{-1})\;. 
\end{eqnarray}
It is then straightforward to extend the action of the $R$-matrix (and its inverse) 
to any function on $SU(2)^{\otimes 2}$.

Note the important property that the symmetric intertwiner is invariant under braidings, i.e.:
\begin{eqnarray}
\iota_s([p];[q]) \; R^{\varepsilon}_{p_i,p_{i+1}} \; = \; R^{\varepsilon}_{q_j,q_{j+1}} \; \iota_s([p];[q]) \; = \; \iota_s([p];[q]),
\end{eqnarray}
for any $i \in [1,a-1]$, $j \in [1,b-1]$ (using the notations of the previous sections) and $\varepsilon \in \{+1,-1\}$.
This is known as the pivotal symmetry of the BC intertwiner.

\subsubsection*{3.2. Particles on the Sphere $S^2$}
We are going to show how the Drinfeld double appears in the context of 3D LQG coupled to point particles.
This is in fact immediate and we have the following theorem:
\begin{theorem}[The Drinfeld double in LQG]\label{doubleinLQG}
Let $[m]=(m_0,\cdots,m_{n-1})$ the masses of $n$ particles; $[m]$ labels also a family of simple representations of $DSU(2)$.
Let us define the trivial map between the partial physical Hilbert space ${\cal H}^{0,n}_{Pphys}([m];G)$ and the tensor product 
$\bigotimes_{i=0}^{n-1} {\cal V}_{\overline{m_i}} \simeq Fun((S^2)^{\times n})$ of simple representations as follows:
\begin{eqnarray}
{\cal F} : {\cal H}^{0,n}_{Pphys}([m];G)  \longrightarrow  \bigotimes_{i=0}^{n-1} {\cal V}_{\overline{m_i}} \;,\;\;\;
\varphi  \longmapsto  {\cal F}(\varphi) :  (\tilde{x}_{0},\cdots,\tilde{x}_{n-1}) \mapsto \mathbb I(\tilde{x}_0)
\varphi(\tilde{x}_1,\cdots,\tilde{x}_{n-1})\;. 
\end{eqnarray}
This map is obviously linear and is trivially extended to the algebra of functions.
Moreover, the partial physical Hilbert space between two states $\varphi$ and $\psi$ is given by:
\begin{eqnarray}
<\varphi;\psi>_{Pphys} \; = \; \iota_s([\overline{m}];0) \vert {\cal F}(\overline{\varphi}\psi)> \;.
\end{eqnarray}
\end{theorem}

{\it Proof.}

This theorem is an immediate consequence of the results of the previous section. $\Box$

\medskip

As a consequence, a (partial) physical state is a tensor product of vectors of simple representations of the quantum double
$DSU(2)$: each particle is represented by a vector of a simple representation. Therefore, the quantum double is the symmetry group
of the quantum theory: given an element $x\otimes f \in DSU(2)$, $x$ is a rotation and $f$ is a translation acting on a one-particle
state. It is immediate to understand that the $SU(2)$ part of the double generates rotations; to see that the $F(SU(2))$ part
generates translations, we take $\vec{a} \in \mathbb R^3$ and we consider the element $f_{\vec{a}}$ defined by
$f_{\vec{a}}(x)=e^{i\vec{p}(x) \cdot \vec{a}}$ where $\vec{p}(x)=m\Lambda \vec{n}$ with $x=\Lambda h(m) \Lambda^{-1}$ ($\vec{n}$
a unitary vector). It is clear that $f_{\vec{a}}$ acts on a state by translation.

\medskip

Let us discuss some consequences of this theorem.

First of all, we note that physical processes conserves the momenta of each particle: a pure momenta state has a non-trivial
physical amplitude with the same pure momenta state.

Next, we say some words concerning computation of expectation value of operators. 
An operator can be viewed as (a product of) the evaluation (on simple representations) of elements of $DSU(2)$: such elements act
naturally on ${\cal H}_{Pphys}^{0,n}([m];G)$. Particular examples of operators are the multiplicative $m_i(f)$ and the derivative operator
$d_i(g)$ respectively associated to a function $f$ and a group element $x$ acting on a given particle $i \in [1,n-1]$ as follows:
\begin{eqnarray} 
m_i(f) \vert \varphi > \; = \; f(\Lambda_i)\varphi(\Lambda_1,\cdots,\Lambda_{n-1}) \;\;\;\;\; \text{and} \;\;\;\;\;
d_i(x)  \vert \varphi > \; = \; \varphi(\Lambda_1,\cdots, x\Lambda_i,\cdots, \Lambda_{n-1})\;.
\end{eqnarray}
Note that the multiplicative operator can be naturally extended to the case where $f$ is a distribution.
In fact, any operator (in our sense) on the partial physical Hilbert space can be decomposed into a sum of products of these basic
operators. We introduce a star (antilinear involutive) operator in the algebra of operator generated by $m_i(f)$ and $d_i(x)$ as follows:
$d_i(x)^\star = d_i(x^{-1})$ and $m_i(f)^\star = m_i(\overline{f})$.

An operator $\cal O$ is unitary in the partial physical Hilbert space if it satisfies the identity
\begin{eqnarray}
<\varphi \vert {\cal O} \psi> \; = \; <{\cal O}^\star \varphi \vert \psi>
\end{eqnarray}
for any state $\varphi$ and $\psi$.
Note that $m_i$ is trivially an unitary operator in the partial physical Hilbert space whereas $d_i$ is not in general.
Both operators are unitary in the partial kinematical Hilbert space. To construct (non trivial) unitary operator in the
partial physical Hilbert space, one needs to consider braiding operators that are a particular combinaison of multiplicative
and derivative operators. We will see their definition in the sequel.

There is a natural adjoint action of $DSU(2)$ on the set of operators defined by:
\begin{eqnarray}\label{actionofD}
X \in DSU(2) \;,\;\;\;\; X \rhd {\cal O} \; = \; S(X_{(1)}) {\cal O} X_{(2)}\;,
\end{eqnarray}
where we have used the Sweedler notation and $\cal O$ is an operator. This action is well-defined.
Among the operators, one distinguishes particularly the so-called observables, i.e.the operators that are invariant under the 
action of the quantum double (\ref{actionofD}). 

Given a physical process involving in and out states denoted $\varphi_{in}$
and $\varphi_{out}$, the expectation value of any operator $\cal O$ during this process is given by:
\begin{eqnarray}
<\varphi_{out}\vert {\cal O} \vert \varphi_{in}>_{Pphys} \; = \; 
\iota([\overline{m}];0)  \vert {\cal F}(\overline{\varphi_{out}} {\cal O} \varphi_{in})>\;.
\end{eqnarray}
The mean value  of any operator $\cal O$ is trivially given by:
\begin{eqnarray}
 <{\cal O}> \; \equiv \; \frac{\text{Tr}({\cal O})}{\text{Tr}(1)} \; = \;
\frac{\iota_s([\overline{m}];0) \vert {\cal F}({\cal O}) >}{\text{Tr} (1)} \;.
\end{eqnarray}
Mean values of any operator are therefore given by coefficients of simple intertwining coefficients. 

Interesting operators are the monodromy around the particles and their trace are quite easy to compute. 
Given a finite dimensional representation $K$ of $SU(2)$ and a loop $\ell$ around
$p$ particles among the $n-1$ particles (let us denote by $\alpha_\ell$ the set of particles), 
a monodromy $M_K(\ell)$ is defined by $M_K(\ell)=\chi_K(H_\ell)$ where the expression of the holonomy $H_\ell$
depends on the choice of the loop $\ell$. If $\ell_i$ is a loop around the particle $i$, then $M_K(\ell_i)$ is obviously
a diagonal operator whose mean value is given by the character $<M_K(\ell_i)> = \chi_K(m_i)$. In the case where $\ell$
is any loop, the computation of the mean value is more involved but the result does not depend on the homotopy class of
the loop $\ell$ and is given by:
\begin{eqnarray}
\text{Tr}(M_K(\ell)) \; = \; \sum_{I,J} Y(I,J,K) \; d_I \; d_J \; \prod_{i \notin \alpha_{\ell}} \frac{\chi_I(h(m_i))}{d_I} 
\prod_{j \in \alpha_{\ell}} \frac{ \chi_J(h(m_j))}{d_J} \;
\end{eqnarray}
where $Y(I,J,K) = 1$ if $I,J,K$ satisfy triangular inequalities and $Y(I,J,K) =0$ if not.

\begin{figure}[h]
\psfrag{I1}{$I_1$}
\psfrag{I2}{$I_2$}
\psfrag{I3}{$I_3$}
\psfrag{In}{$I_{n-1}$}
\psfrag{J}{$J$}
\psfrag{a1}{$a_1$}
\psfrag{a2}{$a_2$}
\psfrag{a3}{$a_3$}
\psfrag{an}{$a_{n-1}$}
\psfrag{m1}{$m_1$}
\psfrag{m2}{$m_2$}
\psfrag{m3}{$m_3$}
\psfrag{mn}{$m_{n-1}$}
\psfrag{m0}{$m_0$}
\psfrag{0}{$0$}
\psfrag{...}{$\cdots$}
\centering
\includegraphics[scale=0.9]{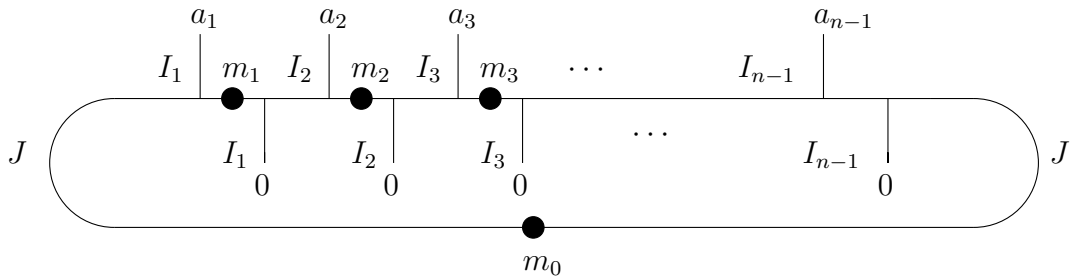}
\caption{\it Pictorial illustration of the expression (\ref{vev}). The dot colored with the mass $m$ denotes insertion of the 
element $h(m)$; any trivalent vertex denotes $(3j)$ coefficients of $SU(2)$; each edge is colored with an irrep of $SU(2)$ and
can end up with a magnetic number like $a_i$ or $0$.}
\label{simplecoefficient}
\end{figure}

One can compute the mean value of more general operators if one knows explicitely simple intertwining coefficients. 
The expression of a simple
intertwining coefficient in the basis of functions $\stackrel{I}{\pi}{}\!^i_j$  
is explicitly given by the formula:
\begin{eqnarray}\label{vev}
<\bigotimes_{i=1}^{n-1}\stackrel{I_i}{\pi}\!\!^{a_i}_{0}>_{Pphys} \; = \; \sum_J d_J \sum_{j_i,k_i} \prod_{i=0}^n 
\stackrel{I_iJ}{C}{}\!^{a_i j_i}_{k_i} 
\end{eqnarray}
in term of the following coefficients:
\begin{eqnarray}
\stackrel{I_iJ}{C}{}^{a_i j_i}_{k_i} \; \equiv \;
\sum_k \stackrel{J}{\pi}{}\!\!^k_k (h(m_i)) \int d\Lambda \stackrel{I_i}{\pi}{}\!\!^a_0(\Lambda) \;
\stackrel{J}{\pi}{}\!\!^{j_i}_k(\Lambda) \; \overline{\stackrel{J}{\pi}{}\!\!^{k_i}_k(\Lambda)} \;,
\end{eqnarray}
which can easily expressed in term of the coefficients of $(3j)$ symbols of $SU(2)$ (see figure (\ref{simplecoefficient}) for 
an illustration).

\medskip

Now, let us discuss some aspects concerning the braiding. First of all, we remark that the $R$ matrix, as introduced 
in the previous section (\ref{Rmatrix}), defines an isomorphism of (partial physical) Hilbert spaces. Indeed, if one considers the
particle labelled by $i \in [1,n-2]$, we define the following map:
\begin{eqnarray}\label{Braidoperator}
R_{i,i+1}: {\cal H}_{Pphys}^{0,n}([m];G) \; \longrightarrow \; {\cal H}_{Pphys}^{0,n}(P_i[m];G)
\end{eqnarray}
where $P_i[m]=(m_0,\cdots, m_{i+1},m_i, \cdots, m_{n-1})$ permutes the particles $i$ and $i+1$. The fact that the previous map is an
isometry is a consequence of the invariance of the simple intertwiner under braidings.
Let us remark that the operator (\ref{Braidoperator}) can be written in term of multiplicative and derivative
basic operators as follows: $$R_{i,i+1}= \int dx \; m_i(\delta_x) \; d_{i+1} (xh(m_i)x^{-1})\;.$$
Then, it is easy to compute the action of the star operator on $R$ and we show that it is an unitary operator.

\begin{figure}[h]
\psfrag{R}{$R_{i}$}
\psfrag{R-}{$R_i^{-1}$}
\psfrag{i}{$i$}
\psfrag{j}{$i+1$}
\psfrag{0}{$0$}
\psfrag{n-1}{$n-1$}
\psfrag{-}{$\cdots$}
\centering
\includegraphics[scale=0.7]{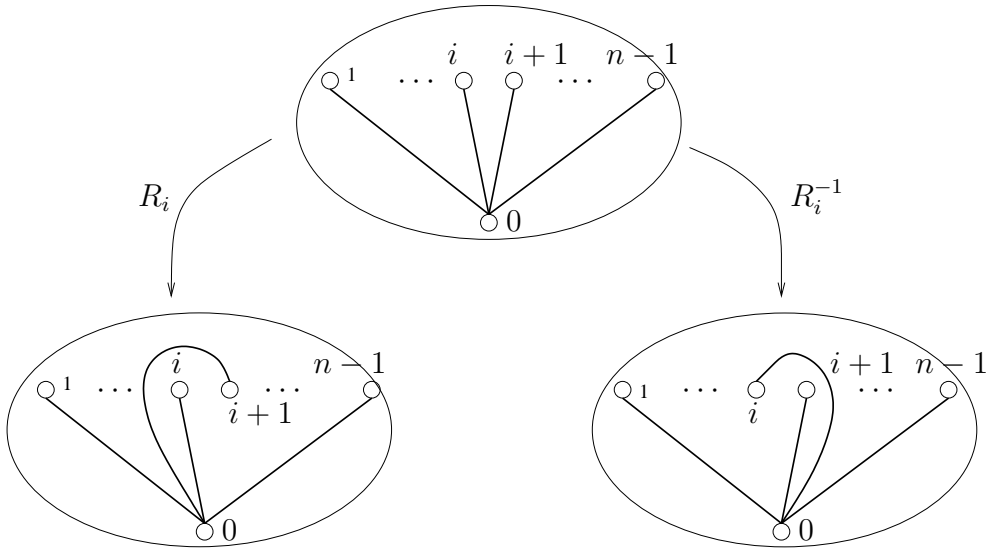}
\caption{\it Illustration of the braiding between two particles $i$ and $i+1$. On the left, the braiding is  associated to the $R$ matrix
whereas it is associated to its inverse $R^{-1}$ on the right.}  
\label{Braid}
\end{figure}

Physically, the braiding operator (\ref{Braidoperator}) corresponds to a braiding between the two spin-network edges that link the 
observer with the particles $i$ and $i+1$. In fact, there is two ways to braid the edges: 
one is associated to the $R$ matrix and the other to its inverse. This point is illustrated in the picture (\ref{Braid}).

This braiding property means that the partial physical Hilbert space does not depend on the order on the particles. 

One can generalize the braiding operator to any pair of particles $(i,j)$ (let us assume that $i<j$). 
There are many ways to do so and a simple example of such an operator is given by:
\begin{eqnarray}\label{generalbraid}
R_{i,j} \; \equiv \; \prod_{k=i}^{j-1} R_{k,k+1} \prod_{\ell = j-2}^i R_{\ell,\ell+1} \;.
\end{eqnarray}
One can as well defines an operator which involves not only $R$ matrices but also its inverse.
The operator (\ref{generalbraid}) defines an isometry between the Hilbert spaces ${\cal H}^{0,n}_{Pphys}([m],G)$ and 
${\cal H}^{0,n}_{Pphys}(P_{i,j}[m],G)$ where $P_{ij}$ permutes the particles $i$ and $j$.

\subsubsection*{3.3. Particles on any Riemann surface}
The generalization of the theorem (\ref{doubleinLQG}) to the case where $\Sigma$ is any Riemann surface (of genus $g$) is quite simple.
To do so, we first need the following proposition:
\begin{proposition}[Reduction of the genus]
Let $p$ be a simple representation. We define the map $\Upsilon_p$ as follows:
\begin{eqnarray}
\Upsilon_p : {\cal H}_{Pphys}^{g,n}([m];G) & \longrightarrow & {\cal H}_{Pphys}^{g-1,n+2}([m],p,\overline{p};G) \\
\psi & \longmapsto & \Upsilon_p(\psi)(A^{g-1,n},\tilde{x},\tilde{y}) = \int da_1 \; db_1 \; \delta(b_1 y h(p) y^{-1}) \; \delta(ayx^{-1})
\; \psi(A^{g,n})  \nonumber
\end{eqnarray}
where we have used the notations of the proposition (\ref{Inclusions}). 

This map satisfies the following property:
\begin{eqnarray}\label{Upsilon}
<\psi,\varphi>_{Pphys} \; = \; \int d\mu(p) \; <\Upsilon_p(\psi),\Upsilon_p(\varphi)>_{Pphys}
\end{eqnarray}
where the partial physical scalar products are respectively those of ${\cal H}_{Pphys}^{g,n}([m];G)$ and 
${\cal H}_{Pphys}^{g-1,n+2}([m],p,\overline{p};G)$. 
\end{proposition}

{\it Proof.}

Given two states $\psi$ and $\varphi$ in ${\cal H}_{Pphys}^{g,n}([m];G)$, 
we start by computing the partial physical scalar product in ${\cal H}_{Pphys}^{g-1,n+2}([m],p,\overline{p};G)$
and a direct calculation shows that:
\begin{eqnarray}
<\Upsilon_p(\psi),\Upsilon_p(\varphi)>_{Pphys} \!\!& = & \!\!\int \prod_{i=1}^n d\tilde{z}_i\prod_{j=2}^g da_jdb_j dx dy 
\overline{\psi(A^{g-1,n},x,yh(\overline{p})y^{-1})}  \varphi(A^{g-1,n},x,yh(\overline{p})y^{-1}))\nonumber \\
&& \delta(h_0 \prod_{i=1}^n z_i h(m_i) z_i^{-1} xyh(p)y^{-1}x^{-1} yh(\overline{p})y^{-1} \prod_{j=2}^g [a_j,b_j]) 
\end{eqnarray}
A straightforward application of the Kirillov formula, $\int dx = \int d\mu(p) d\tilde{\Lambda}$ where $x=\Lambda h(p)\Lambda^{-1}$, 
leads to (\ref{Upsilon}). $\Box$

\medskip

An immediate consequence of this proposition is that one can reduce the case of a Riemann surface to the case of the sphere
(regarding the computation of the partial physical scalar product) and we have the following theorem:
\begin{theorem}[Reduction to the case of the sphere]\label{theo2}
Let $[p]=(p_1,\cdots,p_g)$ a family of $g$ simple representations. We define the map $\Upsilon_{[p]}=\Upsilon_{p_g} \circ 
\cdots \circ \Upsilon_{p_1}$ and the partial physical scalar product between two states $\varphi, \psi \in {\cal H}_{Pphys}^{g,n}([m];G)$
is expressed in term of symmetric intertwiner as follows:
\begin{eqnarray}
<\varphi,\psi>_{Pphys} \; = \; \int \prod_{i=1}^g d\mu(p_i) \; 
\iota_s([m],[p,\overline{p}];0) \vert {\cal F}\circ \Upsilon_{[p]}(\overline{\varphi} \psi)>
\end{eqnarray}
where $[p,\overline{p}]=(p_1,\overline{p}_1,\cdots,p_g,\overline{p}_g)$
and ${\cal F}$ has been introduced in the theorem (\ref{doubleinLQG}).
\end{theorem}

{\it Proof.}

Let $\varphi, \psi \in {\cal H}_{Pphys}^{g,n}([m];G)$.
By a recursive use of the proposition (\ref{Upsilon}), we immediately show that:
\begin{eqnarray}
<\varphi,\psi>_{Pphys} \; = \; \int \prod_{i=1}^g d\mu(p_i) <\Upsilon_{[p]}(\varphi),\Upsilon_{[p]}(\psi)>
\end{eqnarray}
where the scalar product on the r.h.s. is the partial physical scalar product in the case of the sphere.
Then, we conclude immediately thanks to the theorem (\ref{doubleinLQG}). $\Box$

\medskip

Let us discuss some consequences of this theorem. 

1. First of all, the quantum group structure still holds when $\Sigma$ is any Riemann surface. However, to make the contact between
the Hilbert structure and the simple intertwiner is more involved than the case of the sphere. Anyway, the quantum double is
manisfestly the symmetry group of the quantum theory.

\medskip

\begin{figure}[h]
\psfrag{p1}{${\cal P}_1$}
\psfrag{p2}{${\cal P}_2$}
\psfrag{pn-1}{${\cal P}_{n-1}$}
\psfrag{p0}{${\cal P}_0$}
\psfrag{h1}{$\!\!h_1$}
\psfrag{hg}{$h_g$}
\centering
\includegraphics[scale=0.5]{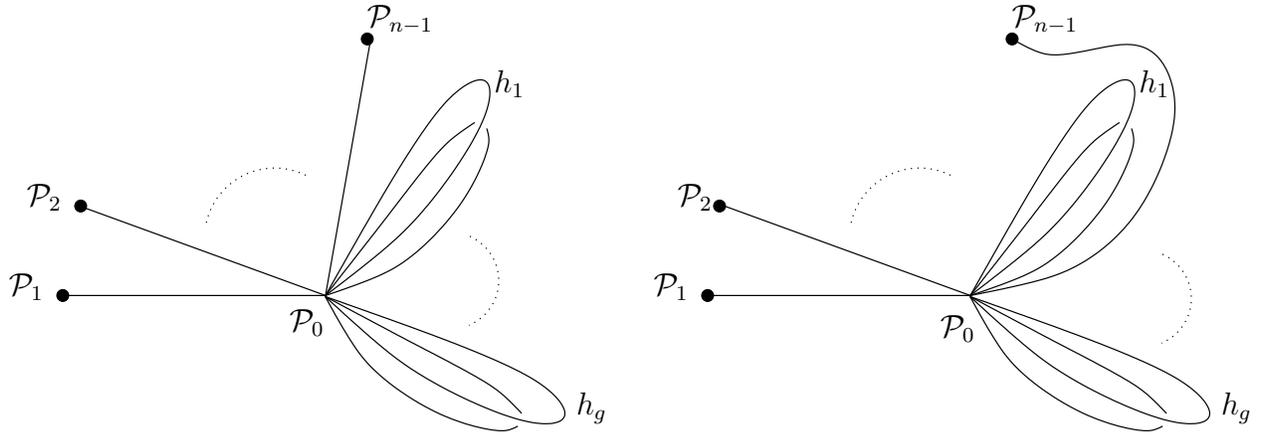}
\caption{\it Pictorial illustration of the braiding: the partial kinematical Hilbert spaces associated to the minimal graphes
on the left and on the right are isometric and the isometry is defined by a ``general'' braiding operator. Note that ${\cal P}_i$
denotes the particles and $h_i$ the handles.}
\label{Braidgeneral}
\end{figure}

2. What about the braiding? We have chosen an order to define the partial physical Hilbert space: this order is illustrated in the picture
(\ref{Braidgeneral}) and consists in putting the particles variables first and the handledodies variables next. 

We have shown previously that permuting two particles variables corresponds to acting with a $R$ matrix. In fact the same thing
happens when one permutes particles with handles or handles with handles. Indeed, in the theorem (\ref{theo2}), we see that one
handle is decomposed into a pair of particles whose masses are $m$ and $\overline{m}$; therefore permuting one particle with one handle
corresponds to permuting this particle with the two particles associated to the handle, as a consequence this operation corresponds
to acting twice with a $R$ matrix.
To be more concrete, let us consider the example
of the picture (\ref{Braidgeneral}). The Hilbert spaces ${\cal H}_{Pphys}^{g,n}([m],\gamma_1;G)$ and 
${\cal H}_{Pphys}^{g,n}([m],\gamma_2;G)$, where $\gamma_1$ and $\gamma_2$ are the minimal graphes associated, are related by the following
isometry:
\begin{eqnarray}
R_{\gamma_1,\gamma_2} \; \equiv \; \int d\mu(p_1) \; \stackrel{m_{n-1},\overline{p}_1}{R}\!\!\!\!\!\!_{n,n+1} \; \circ \;
\stackrel{m_{n-1},{p}_1}{R}\!\!\!\!\!\!_{n-1,n} \; \circ \; \Upsilon_{p_1}  
\end{eqnarray}
where the $R$ matrix $R_{i,i+1}$ acts on the particles variables $i$ and $i+1$. We have explicitely shown the representation 
for the evaluation of the R matrices in order to be clear. On can easily show that this map is an isometry. It is easy to generalize this 
map to the case where $\gamma_2$ is any minimal graph.

\medskip

\begin{figure}[h]
\psfrag{p}{$p$}
\psfrag{pb}{$\overline{p}$}
\psfrag{=}{$= \int d\mu(p)$}
\centering
\includegraphics[scale=0.9]{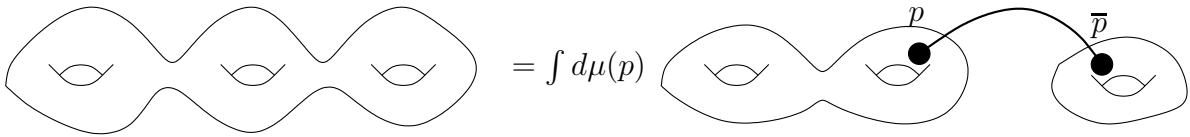}
\caption{\it Illustration of the factorization property of the partial physical scalar product: the physical scalar product on any
Riemann surface can be factorized into partial physical scalar product on the torus.}
\label{factorization}
\end{figure}

3. The next remark concerns some factorization properties of the partial physical scalar product. To clarify what we mean by that
we use the notation $<;>_{Pphys}^{g,n}[m]$ for the partial physical scalar product within this remark and we have:
\begin{eqnarray}
<;>_{Pphys}^{g,n}([m]) \; = \; \int d\mu(p) \; <;>_{Pphys}^{g-1,n+1}([m],p) \; \otimes \; <;>_{Pphys}^{1,1}(\overline{p}) \;. 
\end{eqnarray}
Therefore, the physical scalar product can be factorized into scalar product on the torus.
This property is well-known and is illustrated in the picture of the figure (\ref{factorization}).

\medskip

4. The last remark concerns the evaluation of geometric observables. Indeed, in the case of a non-trivial topology, there exists 
non-trivial ``geometric'' operators which corresponds to monodromies around handles. For instance, the trace of $M_K(a_i)$
where $a_i$ is a loop around one handle is given by:
\begin{eqnarray}
\text{Tr}(M_K(a_i)) \; = \; \sum_J Y(J,J,K) \; \frac{\prod_{i=0}^{n-1} \chi_J(h(m_i))}{d_J^{2g+n-2}} \; = \;
\sum_{J \geq [K/2]} \frac{\prod_{i=0}^{n-1} \chi_J(h(m_i))}{d_J^{2g+n-2}} 
\end{eqnarray}
in the case of a genus $g$ surface with $n$ particles. $[x]$ denotes the entire part of $x$. There is at least two different ways
to proove this identity: one can use the last theorem (\ref{theo2}) to come back to the case of the sphere or one can make a direct
calculation.

\medskip

As a consequence, we have now a complete description \`a la LQG of three dimensional gravity coupled to massive particles.

\section*{4. Conclusion and Perspectives}
In this paper, we have performed the hamiltonian quantization of three dimensional riemannian gravity (with no cosmological
constant) coupled with a finite number of massive spinless point particles using Loop Quantum Gravity techniques.
The work we have proposed contains: a complete description of the (partial) physical Hilbert space in the general case of any 
Riemann surface (i.e. states and scalar product) and a discussion concerning the question of the observables (i.e. their definition
and the computation of their expectation values in some examples). 
We have emphasized the crucial role of the Drinfeld double which is clearly the symmetry group of the quantum theory, as expected.
In the LQG approach, the Drinfeld double appears as a result of the dynamics in the sense that its role becomes obvious when one imposes
the three dimensional analog of the ``hamiltonian'' constraint. This point of view is quite interesting because 
it is a non-trivial 
example where one sees the emergence of a non-compact quantum group (at the level of the physical Hibert space) starting from a 
classical compact group (at the level of the kinematical Hilbert space). Our construction is in that sense very different, at the conceptual level, from the combinatorial quantization, the group field theory quantization or
the spin-foam quantization where the quantum group is put by hand as the basic starting block of these methods.
From the point of view of a particles physicist, our approach might seem more satisfying for we describe particles as simple representations of the "Poincare" (in Riemannian space) group and we notice that, using LQG techniques,
scattering amplitudes of these particles in a quantum background are described in terms of intertwiners of a quantum group. In fact, in three dimensions, we see that the hamiltonian constraint (in the presence of particles) is in fact a simple intertwiners!

Thus, physical states are constructed from simple representations
of the Drinfeld double and the physical scalar product is in fact given by a symmetric (or Barrett-Crane -BC) intertwiner between the
simple representations defining the states. The ressemblance with the usual BC model is  
obvious and asks the question whether the BC model is really a model of four dimensional quantum gravity or a model of three dimensional
gravity coupled to particles. 

\medskip

This paper opens new insights in LQG techniques. The first one would be to see the emergence of the quantum group $SU_q(2)$ when one
imposes the Hamiltonian constraint with a cosmological constant (the quantum parameter $q$ is related to $\Lambda$ in the usual way).
We are currently working in that direction. It seems also possible to generalize our construction to the lorentzian case because we 
finally do not need to make sense of the whole space of cylindrical functions but only of the space of cylindrical functions on one graph,
the minimal graph. Therefore, the obstruction raised in \cite{noncompactLQG} and in \cite{Willis} should not be problematic in our context.
Going from one minimal graph to another reduces to a non-trivial braiding which is still well defined in the Lorentzian 
regime. Moreover, 
one could introduce new types of particles: lightlike, spacelike or timelike particles. We could also think of introducing 
particles that would describe a black hole in the presence of a negative cosmological constant.

\medskip

Another issue to be solved is to extend this construction to the case of spinning particles. This case has been
considered briefly in the spin-foam approach by \cite{Freidel}. In the LQG point of view,
the description of the kinematical Hilbert space has been performed completely in \cite{Noui2} but the way that
the physical scalar product is related to the quantum double remains to be understood. We expect that this link works in the same way as the non-spinning case with the difference that the physical scalar product (or spinning particles scattering amplitudes in a quantum background) is no longer given by a simple intertwiner but a more general $DSU(2)$ intertwiner. In our point of view, the case of spinning particles is not a conceptual issue but only a technical issue. 

\medskip

In fact, the original motivation for this work, is the construction of a self-gravitating
quantum field theory in the haniltonian framework. This means a precise description of the Fock space associated to its creation and
annihilation operators. We are finishing a work in that direction \cite{Fockspace} 
and we have shown in particular that the quantum field theory so
obtained is closely related to the one constructed in \cite{Freidel} in the context of spin-foam models.

\medskip

Finally, the last but not the least is to make use of this construction to describe a coupling of quantum gravity to matter field
in four dimensions. It seems quite obvious how to generalize, at the level of the kinematical Hilbert space, the coupling of spin-networks
to some matter fields. The more interesting question would be to understand what kind of matter that represents and what is the dynamics
of such a coupling (see \cite{BaezPerez} for ideas in that direction).

\subsubsection*{Aknowledgments}
I would like to greatly thank A. Perez. I would also like to thank D. Oriti, J. Ryan and K. Krasnov for stimulating and interesting discussions. I would also like to thank L. Freidel and P. Roche for many many discussions concerning quantum groups and the quantum double.

This work was partially supported by the ANR-06-BLAN-0050-03 LQG-2006.

\subsection*{Appendix A: The Drinfeld double}
We recall the general definition
of the quantum double that we illustrate in the case of a finite group.

\medskip

Let $A$ be a Hopf algebra, the quantum double of $A$ is denoted $D(A)$ (or $DA$), is a quasitriangular Hopf algebra containing $A$
as a sub-Hopf algebra and has different equivalent definitions.

If $A$ is a Hopf algebra, we denote $m$ the multiplication and $\Delta$ the coproduct; it will be useful to introduce a permutation
operator $\sigma$. From $A$, we can construct different Hopf algebra:
$A_{op}$ is the Hopf algebra with multiplication $m_{op}=m\circ \sigma$ and coproduct $\Delta$;
$A^{op}$ is the Hopf algebra with multiplication $m$ and coproduct $\sigma \circ \Delta$;
the dual Hopf algebra $A^{\star}$ construted from the duality bracket $<,>$ between $A$ and $A*$.

\medskip

The quantum double (or Drinfeld double) $D(A)$ is the Hopf algebra defined by the following:
\begin{enumerate}
\item $D(A)=A \otimes A^{\star op}$ as a coalgebra;
\item The algebra law is given by:
\begin{eqnarray}
(x\otimes \xi)(y \otimes \eta) \; = \; \sum_{(y),(\xi)} xy_{(2)} \otimes \xi_{(2)}\eta <\xi_{(1)},S^{-1}y_{(1)}><\xi_{(3)},y_{(3)}>
\end{eqnarray}
where $\Delta(x)=\sum_{(x)}x_{(1)}\otimes x_{(2)}$ and $\Delta_{A^{\star op}}(\xi)=\sum_{xi}\xi_{(1)}
\otimes \xi_{(2)}$ is the usual Sweedler notation.
\end{enumerate}
$D(A)$ is quasitriangular  and the $R$-matrix is given by $R=\sum_i e_i\otimes 1 \otimes 1 \otimes e^i$ where $(e_i)_i$ (resp. $(e^{i})_i$)
is a basis of $A$ (resp. $A^{\star}$).

\medskip

Let us illustrate this construction in 
the case of a finite group to be more concrete. Let $G$ be a finite group and we assume that $A$ is the group algebra of $G$
i.e. $A=\mathbb C[G]$. As a result $A^{\star}=F(G)$, the algebra of functions on $G$, and the quantum double $D(\mathbb C[G])=\mathbb C[G]
\otimes F(G)^{op}$ is usually denoted $D(G)$. A basis of $D(G)$ is $(x\otimes \delta_g)_{x,g\in G}$ and the Hopf algebra 
and coalgebra structures are respectively given by:
\begin{eqnarray}
(x\otimes \delta_g)\cdot(y\otimes \delta_h) \; = \; xy \otimes \delta_h(y^{-1}gy)\;\delta_h \;\;
\Delta(x\otimes \delta_g) = \sum_{g_1,g_2 \vert g_1g_2=H} (x\otimes g_2) \otimes (x\otimes g_1) \;.
\end{eqnarray}
The action of the antipode on the previous basis is given by $S(x\otimes \delta_g)=x^{-1} \otimes \delta_{x^{-1}g^{-1}x}$
and the co-unit is defined by $\epsilon (x\otimes \delta g)=\delta(g)$. 
The $R$-matrix is given by $R=\sum_{x,g}(x\otimes 1)\otimes(1\otimes \delta_g)$.
This construction can be directly generalized to the case where $G$ is locally compact \cite{KM}. 

Representation theory of $D(G)$ has been done by \cite{DPR} and generalized to the compact group case ($G=SU(2)$) in \cite{KM}.
The main results are recalled in the core of the article. 

\subsection*{Appendix B: Normalization of the physical scalar product}

We introduce the group ${\cal B}^{g,n}\simeq G^{4g+n-1}$ which acts on ${\cal H}_{Pphys}^{g,n}([m];G)$: it acts by right $G$-action
of each variable $\tilde{z}_e$ and by right and left $G$-action on each variables $a_i$ or $b_i$.
Given a partial physical state $s$ and an element $y\in {\cal B}^{g,n}$, we will use the notation $B_y(s)$ for the action of $y$
on $s$.
There exists many couples $(g,n)$ such that the state $s$ can be viewed as
an element of ${\cal H}_{Pphys}^{g,n}([m];G)$ and we denote $(g_{min},n_{min})$ the minimal couple. Then, we define the
invariance group transformations of $s$ by ${\cal B}[s]\equiv {\cal B}^{g_{min},n_{min}}$. To be more concrete, ${\cal B}[s]$
transforms only the variables which appear explicitely in the state $s$. There is a natural action of
${\cal B}[s]$ on ${\cal H}^{n,g}[m]$ and we define the sub-Hilbert space 
$C^{g,n}[s] \subset {\cal H}^{n,g}[m]$ as follows:
\begin{eqnarray}
C^{g,n}[s] \; \equiv \; \{ f \in {\cal H}^{g,n}[m] \; \vert \; \forall \; y \in {\cal B}[s] \;,\; \; B_y(f)=f\}\;.
\end{eqnarray}

\begin{proposition}[Inclusions]\label{Inclusions}
Let $v,u \in {\cal H}_{Pphys}^{g,n}([m];G)$ defined by $u=tr([a_1,b_1])$ and
$v=tr(z_{n-2}h(m_{n-2})z_{n-2}^{-1}z_{n-1}h(m_{n-1})z_{n-1}^{-1})$ where $tr$ denotes the trace
in the fundamental representation of $SU(2)$. We have the following inclusions of Hilbert spaces:
\begin{eqnarray}
C^{g,n}[u]& \hookrightarrow &  \int^{\oplus} \!\! d\mu(p) \; {\cal H}_{Pphys}^{g-1,n+2}([m],p;\overline{p};G) \label{inclu1}\;,\\
C^{g,n}[v]& \hookrightarrow & \int^{\oplus} \!\! d\mu(q) \; {\cal H}_{Pphys}^{g,n-1}(m_0,\cdots,m_{n-3},
\overline{q};G) \otimes {\cal H}_{Pphys}^{0,3}(q,m_{n-2},m_{n-1};G)\;.\label{inclu2}
\end{eqnarray}
The measure are given by $d\mu(p)=\frac{1}{\pi}\sin^2p dp$ where $dp$ is the measure on the circle $S^1$ such that $\int dp=2\pi$.
We have introduced the notation $\overline{p}=2\pi-p$.
\end{proposition}

{\it Proof.}

We concentrate on the first map (\ref{inclu1}) and we show that an explicit map is given as follows:
\begin{eqnarray}
f & \longmapsto & \phi(A^{g-1,n},\tilde{x}_1,\tilde{y}_1) \; \equiv \; \int \!da_1 db_1 \; 
\delta(a_1x_1h(\overline{p})x_1^{-1}) \; \delta(b_1a_1b_1^{-1}{y}_1h({p})y^{-1}_1) \; f(A^{g,n})
\end{eqnarray}
where the family $A^{g,n}=(a_1,b_1;\cdots;a_g,b_g;\tilde{z}_1,\cdots,\tilde{z}_{n-1})$ contains the $n+2g$ arguments of the function $f$
and $A^{g-1,n}=(a_2,b_2;\cdots;a_g,b_g;\tilde{z}_1,\cdots,\tilde{z}_{n-1})$ contains $n+2g-2$ arguments.
The norm of the function $\phi$ viewed as an element of the space ${\cal H}^{n+2,g-1}([m],p,\overline{p};G)$ is given by:
\begin{eqnarray*}
\vert\!\vert \phi \vert \! \vert_{Pphys}^2 & = & \int \prod_{e=1}^{n-1}d\tilde{z}_e \; d\tilde{x}_1d\tilde{y}_1 \; \prod_{i=2}^g da_i db_i 
\vert \phi(A^{g-1,n},\tilde{x}_1,\tilde{y}_1) \vert^2 \\
\;\;\;\;\;\;\;\;\;\;\;\;\;\;\;\;\;\;\;\;\;\;\;\;\;\;\;\;\;\;\;\;\;\;\;
&& \delta(h(m_0) \prod_{e=1}^{n-1}z_e h(m_e) z_e^{-1} x_1 h(p) x_1^{-1}y_1 h(\overline{p}) y_1^{-1} \prod_{i=2}^{g}[a_i,b_i])\;.
\end{eqnarray*}
It is then straightforward to verify that:
\begin{eqnarray*}
\int d\mu(p) \; \vert\!\vert \phi \vert \! \vert_{Pphys}^2 & = & \int \prod_{e=1}^{n-1} d\tilde{z}_e \prod_{i=1}^g da_i db_i \; 
\delta(h(m_0) \prod_{e=1}^{n-1}z_e h(m_e) z_e^{-1} \prod_{i=1}^{g}[a_i,b_i])\\
&& \int dx \; \delta([x,a_1]) \; \overline{f(a_1,b_1;\cdots)} f(a_1,b_1x;\cdots) \; .
\end{eqnarray*}
As $f \in C^{g,n}[u]$ and $B_x(u)=u$ if $[x,a_1]=1$ then we conclude immediately that:
\begin{eqnarray}
\int d\mu(p) \; \vert\!\vert \phi \vert \! \vert_{Pphys}^2 \; = \; \vert\!\vert f \vert \! \vert_{Pphys}^2 \;.
\end{eqnarray}
Therefore, the first inclusion (\ref{inclu1}) is proven.

To proove the second one (\ref{inclu2}), we proceed in the same way.
We start by claiming that the following map trivially realises the injection (\ref{inclu2}):
\begin{eqnarray}
f  \longmapsto \psi((A^{g,n-2},\tilde{x}) \otimes (\tilde{z}_{n-2},\tilde{z}_{n-1})) = f(A^{g,n})  
e^{g,n-2}(m_0,\cdots,m_{n-3},q)\otimes  e^{0,3}(\overline{q},m_{n-2},m_{n-1})\;.
\end{eqnarray}
Note that $\psi$ is viewed as an element of ${\cal H}_{Pphys}^{g,n-1}(m_0,\cdots,m_{n-3},
\overline{q};G) \otimes {\cal H}_{Pphys}^{0,3}(q,m_{n-2},m_{n-1};G)$ and we have introduced the notation
$(A^{g,n-2},x)=(a_1,b_1;\cdots,a_g,b_g;\tilde{z}_1,\cdots,\tilde{z}_{n-3},\tilde{x})$. The norm of $\psi$
is trivially given by:
\begin{eqnarray*}
\vert\!\vert \phi \vert \! \vert_{Pphys}^2 & = & \int \prod_{i=1}^{n-1}d\tilde{z}_e \; d\tilde{x} \;\prod_{i=1}^g da_i db_i \;
\vert f(A^{g,n}) \vert^2 \delta(h(m_0)\prod_{e=1}^{n-3} z_e h(m_e) z_e^{-1} xh(q)x^{-1}\prod_{i=1}^g[a_i,b_i]) \\
& & 
\; \delta(h(\overline{q})z_{n-2} h(m_{n-2}) z_{n-2}^{-1}
z_{n-1} h(m_{n-1}) z_{n-1}^{-1}) \;.
\end{eqnarray*}
Therefore, we verify easily that:
\begin{eqnarray*}
\int d\mu(q)\; \vert\!\vert \phi \vert \! \vert_{Pphys}^2 & = & \int \prod_{i=1}^{n-1}d\tilde{z}_e \; d\tilde{x} \;\prod_{i=1}^g 
da_i db_i \; \vert f(a_1,b_1;\cdots,a_g,b_g;\tilde{z}_1,\cdots,\tilde{z}_{n-3},x\tilde{z}_{n-2},x\tilde{z}_{n-1}) \vert^2 \\
&& \delta(h(m_0)\prod_{i=1}^{n-1}z_e h(m_e) z_e^{-1}\prod_{i=1}^g[a_i,b_i])\;.
\end{eqnarray*}
Finally, as $f \in C^{g,n}[v]$, then we conclude that:
\begin{eqnarray*}
\int d\mu(q)\; \vert\!\vert \phi \vert \! \vert_{Pphys}^2 \; = \; \vert\!\vert f \vert \! \vert_{Pphys}^2 \;,
\end{eqnarray*}
and the Hilbert spaces inclusion (\ref{inclu2}) is proven. $\Box$

\medskip

It is clear that the unit element $e^{g,n}[m] \in C^{n,g}[u]$ for any function $u$. Therefore, we can decompose the unit 
as a tensor product of unit elements of partial Hilbert spaces associated to the sphere with three particles. This is a 
direct consequence of the previous proposition and the explicit decomposition of $e^{g,n}[m]$ is given by:
\begin{eqnarray}\label{decompunit}
e^{g,n}[m] & = & \int \prod_{i=0}^{g-1} d\mu(p_i) \prod_{j=0}^{g-1} d\mu(m_{n+i}) \prod_{k=1}^{n+g-2} d\mu(q_k) \nonumber \\
&&\bigotimes_{a=0}^{g-1} e^{0,3}(p_a,\overline{p_a},m_{n+a}) \; \bigotimes_{b=1}^{n+g-2}e^{0,3}(\overline{r_{b-1}},m_b,r_b)
\end{eqnarray}
with the conventions  $r_0=\overline{m_0}$ and $r_{n+g-2}=m_{n+g-1}$.
As a consequence, one can compute the norm of the identity in two ways and therefore one can relate the coefficients $\lambda(g,n,[m])$
to the coefficients $\lambda(0,3,[m])$. Let us assume that the coefficients $\lambda(0,3,[m])=\lambda$ do not depend on the values
of the masses, then the relation (\ref{decompunit}) implies that:
\begin{eqnarray}
\vert\!\vert e^{g,n}[m] \vert \!\vert^2_{Pphys} & = & \lambda^{n+2g-2} 
\int \prod_{i=0}^{g-1} d\mu(p_i) \prod_{j=0}^{g-1} d\mu(m_{n+i}) \prod_{k=1}^{n+g-2} d\mu(q_k) \;\nonumber \\
&&\prod_{a=0}^{g-1} \vert\!\vert e^{0,3}(p_a,\overline{p_a},m_{n+a}) \vert \!\vert^2_{Pphys}
\prod_{b=1}^{n+g-2}\vert\!\vert e^{0,3}(\overline{r_{b-1}},m_b,r_b)\vert\!\vert ^2_{Pphys}
\end{eqnarray}
After some direct calculations, we show that:
\begin{eqnarray}
\vert\!\vert e^{g,n}[m] \vert \!\vert^2_{Pphys} \; = \; \lambda^{n+2g-2}  \;
\sum_{k=1}^{+\infty}\; {k^{1-2g-n}}\; \prod_{e=0}^{n-1} \frac{\sin(km_e)}
{\sin(m_e)} \; .
\end{eqnarray}
Then, by comparison with the expression (\ref{Unitnorm}), we conclude that $\lambda(g,n.[m])$ is also independent of the
values of the masses and we have:
\begin{eqnarray}
\lambda(g,n,[m]) \; \equiv \;\lambda(g,n) \; = \;\lambda^{n+2g-2}\;.
\end{eqnarray}
This fixes the normalization of the partial physical scalar product.

\bibliographystyle{unsrt}

\end{document}